\documentclass[fleqn,usenatbib]{mnras}

\usepackage[T1]{fontenc}

\DeclareRobustCommand{\VAN}[3]{#2}
\let\VANthebibliography\thebibliography
\def\thebibliography{\DeclareRobustCommand{\VAN}[3]{##3}\VANthebibliography}

\usepackage{graphicx}	
\usepackage{amsmath}	
\usepackage{wasysym}    
\usepackage{verbatim}

\usepackage[dvipsnames]{xcolor}

\usepackage{newtxtext,newtxmath}

\makeatletter 
  \patchcmd{\NAT@citex}
    {\@citea\NAT@hyper@{%
      \NAT@nmfmt{\NAT@nm}%
      \hyper@natlinkbreak{\NAT@aysep\NAT@spacechar}{\@citeb\@extra@b@citeb}%
      \NAT@date}}
    {\@citea\NAT@nmfmt{\NAT@nm}%
    \NAT@aysep\NAT@spacechar\NAT@hyper@{\NAT@date}}{}{}

  \patchcmd{\NAT@citex}
    {\@citea\NAT@hyper@{%
      \NAT@nmfmt{\NAT@nm}%
      \hyper@natlinkbreak{\NAT@spacechar\NAT@@open\if*#1*\else#1\NAT@spacechar\fi}%
        {\@citeb\@extra@b@citeb}%
      \NAT@date}}
    {\@citea\NAT@nmfmt{\NAT@nm}%
    \NAT@spacechar\NAT@@open\if*#1*\else#1\NAT@spacechar\fi\NAT@hyper@{\NAT@date}}
    {}{}
\makeatother

\newcommand{\msun}{M$_{\astrosun}$} 
\newcommand{\mtwo}{$M_{200}$}
\newcommand{\mstr}{$M_\star$}
\newcommand{\rcore}{$r_\text{core}$}
\newcommand{\rh}{$r_{h\star}$}
\newcommand{\rhoth}{$\rho_\text{th}$}
\newcommand{\esf}{$\varepsilon_\text{sf}$}
\newcommand{\vcirc}{$v_\text{circ}$}
\newcommand{\e}[1]{$\times 10^ {#1}$}
\newcommand{\lcdm}{$\Lambda$CDM}
\title[Real and counterfeit cores in \texttt{SMUGGLE}]{Real and counterfeit cores: how feedback expands halos and disrupts tracers of inner gravitational potential in dwarf galaxies}

\author[E.~D.~Jahn et al.]{%
Ethan D. Jahn,$^{1}$\thanks{E-mail: ejahn003@ucr.edu}
Laura V. Sales,$^{1}$
Federico Marinacci,$^{2}$
Mark Vogelsberger,$^{3}$\newauthor
Paul Torrey,$^{4}$
Jia Qi,$^{4}$
Aaron Smith,$^{3}$\thanks{NHFP Einstein Fellow.}
Hui Li,$^{5}$\thanks{NHFP Hubble Fellow.}
Rahul Kannan,$^{6}$
Jan D. Burger$^{7}$\newauthor and
Jes{\'u}s Zavala$^{7}$
\\
$^{1}$Department of Physics \& Astronomy, University of California, Riverside, CA 92507, USA\\
$^{2}$Department of Physics \& Astronomy ``Augusto Righi'', University of Bologna, via Gobetti 93/2, 40129, Bologna, Italy\\
$^{3}$Kavli Institute for Astrophysics and Space Research, Massachusetts Institute of Technology, Cambridge, MA 02139, USA\\
$^{4}$Department of Astronomy, University of Florida, 211 Bryant Space Science Center, Gainesville, FL 32611, USA\\
$^{5}$Department of Astronomy, Columbia University, New York, NY 10027, USA\\
$^{6}$Institute for Theory and Computation, Harvard-Smithsonian Center for Astrophysics, 60 Garden Street, Cambridge, MA 02138, USA\\
$^{7}$Center for Astrophysics and Cosmology, Science Institute, University of Iceland, Dunhagi 5, 107 Reykjavik, Iceland
}

\date{Accepted XXX. Received YYY; in original form ZZZ}

\pubyear{2021}

\begin{document}
\label{firstpage}
\pagerange{\pageref{firstpage}--\pageref{lastpage}}
\maketitle

\begin{abstract}
The tension between the diverging density profiles in Lambda Cold Dark Matter (\lcdm) simulations and the constant-density inner regions of observed galaxies is a long-standing challenge known as the `core-cusp' problem. We demonstrate that the \texttt{SMUGGLE} galaxy formation model implemented in the \textsc{Arepo} moving mesh code forms constant-density cores in idealized dwarf galaxies of \mstr~$\approx$~8\e{7}~\msun~with initially cuspy dark matter halos of \mtwo~$\approx 10^{10}$~\msun. Identical initial conditions run with the \citeauthor{SpringelHernquist2003} (2003; SH03) feedback model preserve cuspiness. Literature on the subject has pointed to the low density threshold for star formation, \rhoth, in SH03-like models as an obstacle to baryon-induced core formation. Using a \texttt{SMUGGLE} run with equal \rhoth~to SH03, we demonstrate that core formation can proceed at low density thresholds, indicating that \rhoth~is insufficient on its own to determine whether a galaxy develops a core. 
We suggest that 
the ability to resolve a multiphase interstellar medium at sufficiently high densities
is a more reliable indicator of core formation than any individual model parameter. 
In \texttt{SMUGGLE}, core formation is accompanied by large degrees of non-circular motion, with gas rotational velocity profiles that consistently fall below the circular velocity $v_\text{circ} = \sqrt{GM/R}$ out to $\sim 2$ kpc. This may artificially mimic larger core sizes when derived from observable quantities compared to the size measured from the dark matter distribution ($\sim 0.5$\,kpc), highlighting the need for careful modeling in the inner regions of dwarfs to infer the true distribution of dark matter. 
\end{abstract}



\section{Introduction}

The difference between the structure of dark matter (DM) halos as predicted by the Lambda-Cold Dark Matter cosmological model (\lcdm) and that which is inferred by observations of gas rotational profiles in galaxies is a long-standing problem in modern cosmology \citep[][]{Moore1994,FloresPrimack1994} with a wide range of postulated solutions. The structure of DM halos as predicted by DM-only simulations \citep[e.g.][]{white1978,Springel2008} is characterized by steeply rising density profiles (`cusps') in the inner regions of halos, parameterized by the NFW profile \citep[][]{navarro1996CDMhalos} which gives a power-law slope $\alpha$ of this inner profile of $-1$. Early measurements of rotation curves in dwarf galaxies have shown regions of constant density known as `cores' with power-law slopes of $\alpha \sim 0$ \citep[e.g.][]{Burkert1995,deblok2001,deBlok2008,KuzioDeNaray2008}. While there is substantial evidence for the existence of cores in dwarf galaxies \citep[e.g.][]{Gilmore2007,Kormendy2009,Oh2011,Oh2015,Lelli2016}, there is debate over the reliability of certain techniques for the inference of the true dark matter potential \citep{Genina2018,Oman2019}.


Another complication to this dilemma is that observed rotation curves in dwarf galaxies exhibit a wide variety of behavior, including rotation curves that rise more rapidly than the NFW profile, consistent with a contraction of the halo, and those that rise significantly more slowly, consistent with expansion. Despite their success in reproducing many observed properties of galaxies, both local and statistical, \citep[][]{Vogelsberger2020}, hydrodynamical simulations of galaxy formation have consistently predicted a uniform shape for rotation curves, posing a problem in replicating the observed diversity \citep{Oman2015,Oman2019,read2016rc,Santos-Santos2018,Santos-Santos2020}.

A theoretically appealing solution to these discrepancies is that the nature of DM is more complex than proposed in \lcdm. Proposed models include warm dark matter \citep{Dodelson1994,Bode2001} and self-interacting dark matter \citep[SIDM,][]{Yoshida2000,Spergel2000,Vogelsberger2012,Vogelsberger2019,Rocha2013,TulinYu2018}. SIDM has been fairly successful in reproducing diverse rotation curves \citep[e.g.][]{Creasey2017,Ren2019,Kaplinghat2020} and explaining the diversity of MW satellites \citep[][]{Zavala2019}. It is worth noting, however, that results for SIDM can depend strongly on the adopted cross-section. Another interesting proposal includes a new hypothetical ultra-light scalar particle with a de Broglie wavelength on astrophysical scales, forming a Bose-Einstein condensate the size of the DM halo, known as fuzzy dark matter \citep[][]{Hu2000,Mocz2017,Lancaster2020,Burkert2020}. While these models prove viable alternatives to \lcdm~with testable predictions \citep[][]{Robles2017,Bozek2019}, they may remain difficult to distinguish from CDM on small scales, especially when the effects of galaxy formation are taken into account \citep{Elbert2018,Fitts2019}. 

It has also been proposed that the feedback-driven motion of baryons within the halo can gravitationally perturb the dark matter potential, leading to expansion \citep{navarro1996}. The repeated outflow of gas following bursts of star formation (SF) has been demonstrated to be a more realistic mechanism for core formation than single, highly violent outbursts \citep[][]{ReadGilmore2005,Governato2010}. This framework was theoretically quantified by \citet{pontzengovernato2012} who introduced an analytical model for core formation in which dark matter particles acquire energy and migrate to more distant orbits via repeated oscillations in the central gravitational potential, driven by supernova (SN) feedback. 

Since the physics of star formation and feedback have not been fully constrained, different effective models of interstellar medium (ISM) physics implemented across the literature have produced different outcomes. For example, the Illustris simulations have been successful in reproducing many properties of galaxies \citep[][]{Genel2014,v14illustris,v14nature}, but have not been able to produce DM cores \citep[][]{Chua2019}. The EAGLE simulations \citep[][]{Schaye2015} have also been shown to not produce DM cores under their fiducial model \citep[][]{Schaller2015,benitezllambay2019}. Zoom-in simulations using the same prescriptions as EAGLE and Illustris have been performed and similarly demonstrate an inability to induce expansion in the DM halo \citep[e.g.][]{bose2019}, indicating that resolution is not responsible for this effect in these models. Meanwhile, other simulations, including \citet{Zolotov2012}, the FIRE project \citep[][]{hopkins2014fire1,hopkins2018fire2,chan2015,wetzel2016LATTE,fitts2017}, and NIHAO \citep[][]{wang2015nihao,tollet2016,Dutton2016}, have been able to produce cores in dwarf galaxies that more closely match observations, indicating that the prediction of DM cores is model-dependent to some degree.


Differences in the modeling of baryonic physics have long been quantified by the SF density threshold, which is the minimum gas density required to form a star particle. \citet{pontzengovernato2012} showed that cosmological zoom-in simulations run with the \textsc{Gasoline} code were unable to induce core formation when using a value of \rhoth\,$=0.1$~cm$^{-3}$, but cores did indeed form when increased to \rhoth\,$=100$~cm$^{-3}$, a value consistent with the observed densities of molecular clouds \citep[][]{Ferriere2001}. Recent work has therefore focused heavily on this parameter, arriving at similar conclusions within the EAGLE simulations \citep[][]{benitezllambay2019}, and NIHAO \citep[][]{Dutton2019}. It has long been reported that `bursty' SF drives repeated outflows, thereby expanding the DM halo by driving mass to the outer regions \citep[][]{Brooks2014}. \citet{benitezllambay2019} conclude from their numerical tests on the density threshold that rapid fluctuations in gas content resulting from bursty SF are insufficient to alter the inner DM halo, but that gas must accrete to high levels of density, dominating the inner gravitational potential before being blown away in order to induce core formation. They also make note that there is no simple relation between SF history and core formation. \citet{Dutton2019} also find that a higher value of \rhoth~induces cores in the NIHAO simulations, but their analysis suggests that fluctuations in SF feedback (and therefore gas content) must occur on \textit{sub-dynamical} timescales in order to induce core formation. Both authors agree that SF burstiness is insufficient to fully explain halo expansion, and that the density threshold is strongly indicative of a resulting flattened inner DM distribution. 

The energetics of core formation discussed in \citet{pontzengovernato2012} require 
rapid motion of sufficiently dense gas clouds in the inner regions of galaxies in order to perturb the gravitational potential and transfer DM to larger orbits.
High resolution simulations that lack detailed physical modelling are unable to capture the small-scale effects of energetic coupling between SF and the ISM due to the use of low star formation threshold, often with \rhoth~$= 0.1 \text{cm}^{-3}$, as well as effective equations of state rather than explicitly implemented cooling physics. Meanwhile, detailed ISM models that self-consistently treat a multiphase, structured ISM are relatively new and have not been directly applied to the problem of core formation. In short, the majority of models that have been used to study this problem are empirically calibrated to reproduce scaling relations of populations of galaxies and implemented in large-volume simulations. These models have been adapted to high resolution zoom-in simulations, with mixed results \citep[][]{benitezllambay2019,bose2019}. Fewer studies have focused on studying core formation as a thoroughly small-scale problem, requiring both high resolution zoom-in simulations and models that capture the local details of physical processes relevant to the state of the ISM. More details of the varying approaches to galaxy modeling are given in a recent review of cosmological simulations \citep[][]{Vogelsberger2020}.

While there is broad agreement in the literature that high thresholds induce cores \citep[e.g.][among the previously listed]{Governato2010,Maccio2012,Teyssier2013,dicintio2014SMHM} and low thresholds do not do not \citep[][]{Oman2015,Schaller2015,bose2019}, there have been limited systematic investigations of the physical outcomes of modeling choices, including comparative analyses of parameters within the same overall modeling scheme. The consistency of models with similar \rhoth~does not rule out the possibility that other modeling choices contribute to halo expansion, including ones that cannot be neatly quantified by a single parameter.

Beyond the physical effects of baryons, difficulties in observing and modeling gas rotation curves in galaxies have led to speculation that large uncertainties might be partially responsible for the observed diversity of galactic rotation curves.  While extensive work has been done to improve observational techniques for estimating velocity profiles \citep[][]{KuzioDeNaray2006,KuzioDeNaray2008,Adams2014}, techniques based on alignment of metallicity populations \citep[e.g.][]{Walker2011} and tilted-ring modeling \citep[e.g.][]{Rogstad1974,Iorio2017} have been recently been demonstrated via application to the APOSTLE simulations to predict DM cores when none actually exist \citep[][]{Genina2018,Oman2019}. This, combined with the large degenerecies in modeling rotational velocities in the presence of non-circular motions \citep[][]{Marasco2018,Santos-Santos2020} suggest that the observed diversity of rotation curves might not be solely a result of physical processes within galaxies, be they baryonic or dark.

In this study, we compare the novel Stars and MUltiphase Gas in GaLaxiEs (\texttt{SMUGGLE}) feedback model \citep[][]{Marinacci2019} to the classic \citeauthor{SpringelHernquist2003} (2003; SH03 hereafter) model, as they represent two paradigms of galaxy formation modeling (i.e. top-down -- SH03, and bottom-up -- \texttt{SMUGGLE}) while implementing the same method of solving gravity+hydrodynamics \citep[\textsc{Arepo,}][]{arepo}. We aim to investigate the differences in and relationship between galaxy formation and DM distribution within these two modeling paradigms in a controlled environment through the use of idealized simulations of a single dwarf galaxy. We also implement variations in model parameters (density threshold and local SH efficiency) within \texttt{SMUGGLE} to shed light on their relevance to core formation within this model, and how their differential effects within this model compare to previous numerical experiments.

The paper is organized as follows: in Section \ref{sec:methods}, we discuss the set up of our isolated dwarf galaxy simulations; in Section \ref{sec:smuggle_and_SH03}, we compare the phenomenological differences between an isolated dwarf galaxy (\mstr~$\sim 10^8$ \msun, \mtwo~$\sim 10^{10}$ \msun) run with each model, and then introduce variations in the \texttt{SMUGGLE} model to investigate the physical nature of core formation in Section \ref{sec:ISM}. We conclude in Section \ref{sec:structure} by examining the morphology of each run, including an investigation of the variation of rotational velocity curves of gas. We summarize our findings in Section \ref{sec:summary}.

\begin{figure*}
    \centering
    \includegraphics[width=0.95\textwidth]{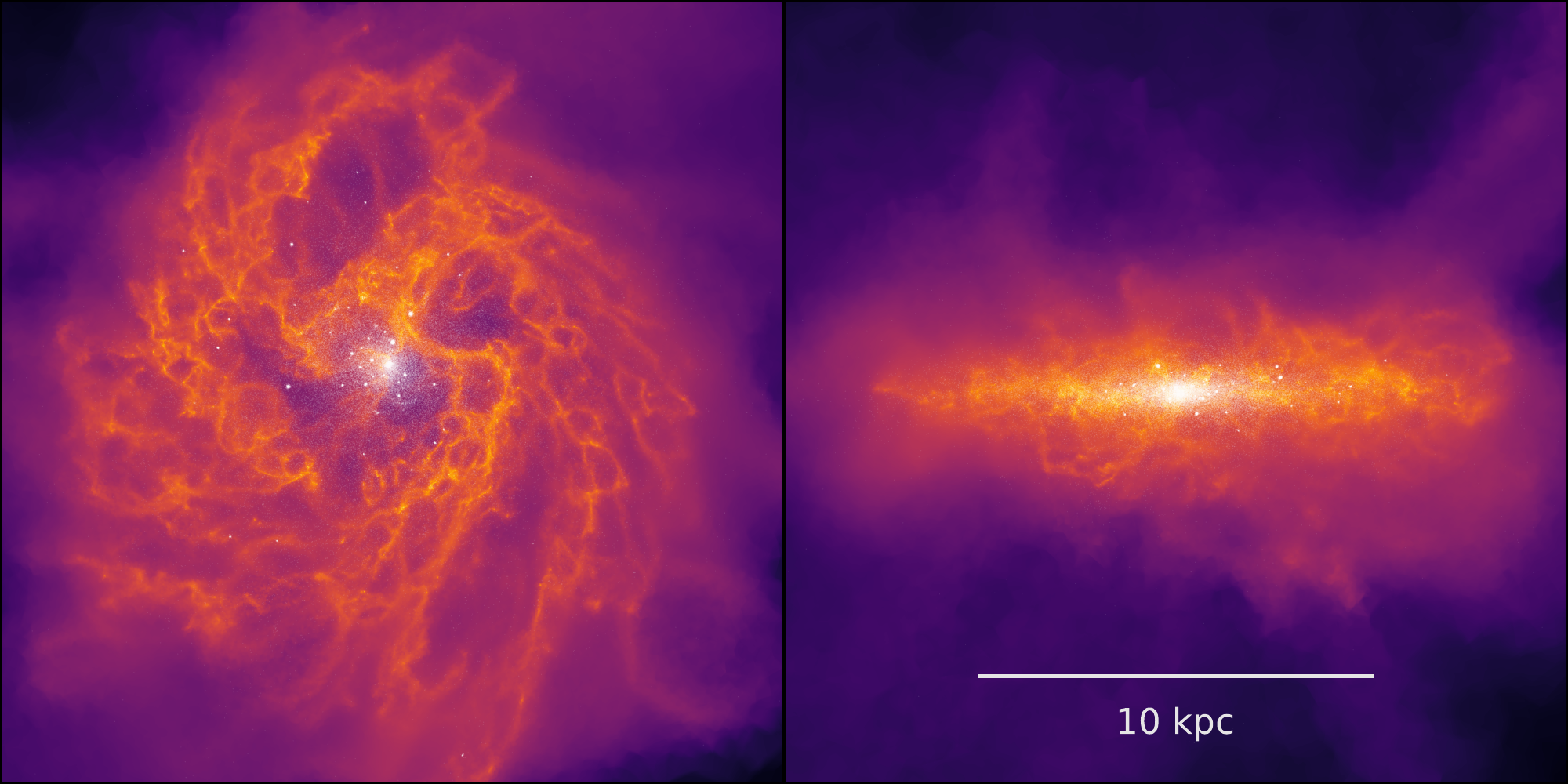}
    \caption{Face-on and edge-on surface density projections of the isolated dwarf galaxy on the fiducial \texttt{SMUGGLE} model generated with the Cosmic Ly$\alpha$ Transfer code \citep[COLT,][]{Smith2015,Smith2017}. Stars formed during the duration of the simulation are shown in white, with gas color-weighted according to surface mass density. The width of each frame is 20 kpc. The richly structured ISM is a result of the detailed ISM physics included in the \texttt{SMUGGLE} model.}
    \label{fig:my_label}
\end{figure*}


\section{Methods}
\label{sec:methods}
We analyze a set of high-resolution, idealized simulations of isolated dwarf galaxies of \mstr~$\approx 10^8$ \msun~in halos of mass \mtwo~$\approx 10^{10}$ \msun~run with the moving mesh code \textsc{Arepo} \citep{arepo,Weinberger2020}. This scale of stellar mass to halo mass has been demonstrated to form feedback-driven cores in other simulation codes \citep[e.g.][]{dicintio2014SMHM,tollet2016}.

Initial conditions were generated via the method described in \citet{Springel2005}, while star formation and feedback were subsequently enabled via the  \texttt{SMUGGLE} model \citep{Marinacci2019}. \texttt{SMUGGLE} implements a wide variety of sub-resolution processes, including gas heating and cooling from which a detailed, multiphase inter-stellar medium (ISM) emerges, a stochastic formation process for stars, and feedback via supernovae (SNe), radiation, and stellar winds. 

Previous work with \texttt{SMUGGLE} includes \citet{Li2020}, who study the formation of giant molecular clouds (GMCs) in Milky-Way mass galaxies, in particular the response of GMCs to various choices of the local star formation efficiency - a parameter we study here as well. They find that \texttt{SMUGGLE} is able to regulate star formation through feedback, with a 3-fold increase in star formation rate (SFR) in runs with no feedback processes enabled. This result is encouraging as it enables self-consistent prediction of kpc-scale galaxy properties. Further, they demonstrate that SN feedback disrupts the spatial correlation of GMCs on scales $> 0.2$ kpc, which is relevant to our discussion on core formation later on. In addition, the \texttt{SMUGGLE} has been further refined with the development of a state-of-the-art model for the treatment of radiation fields, dust physics, molecular chemistry, and metal cooling by \citet{Kannan2020}. This model is able to produce a more complex picture of the ISM of galaxies while maintaining consistent global properties, such as SFR.

\subsection{The \texttt{SMUGGLE} ISM Model}
\label{sec:SMUGGLEmodel}
In this work, we implement the standard \texttt{SMUGGLE} model as described in \citet{Marinacci2019}.
Here we summarize the main physical modeling choices. The primary processes include gravitational hydrodynamics, which is solved by \textsc{Arepo} \citep[][]{arepo}, gas heating and cooling which produce an emergent multiphase ISM, the stochastic formation of star particles, and feedback from stars and SNe. 

\subsubsection{Heating and cooling}
One of the biggest differences in \texttt{SMUGGLE} compared to previously implemented ISM models in {\sc arepo} \citep[e.g. SH03,][]{v14illustris,Pillepich2018} is its ability to explicitly model a cold gas phase with temperatures falling below T$_\text{gas} \sim 10^4$ K. First, a primordial mix of Hydrogen and Helium is modeled by a network of two-body processes including collisions, recombination, Compton cooling via CMB photons \citep[][]{Ikeuchi1986}, and UV-background photoionization \citep[][]{faucher2009}. 

Cooling has two main regimes, metal-line cooling driving the gas temperature down to the warm phase (T$_\text{gas} \sim 10^4$ K) -- which was included in previous ISM models -- while fine structure and molecular cooling implemented in \texttt{SMUGGLE} allows the gas to further cool to $T \sim 10$ K. Cooling rates are calculated in a UV background with the \citet{hopkins2018fire2} fit as a function of temperature, metallicity, gas density, and redshift, with self-shielding taken into account at $z \leq 6$ as in \citet{rahmati2013}. The calculated rates are then scaled to the metallicity of the gas cell. By default, metallicities are updated self-consistently as the simulation evolves in {\sc arepo}. However, for idealized set-ups metallicity can be fixed to offset the lack of replenishment of pristine gas from cosmological infall. For simplicity, in this paper we fix the metallicity of our idealized runs to the solar value. 

\subsubsection{Star formation}
\label{sec:starformation}

Star particles representing single stellar populations with a \citet{chabrier2001} initial mass function are formed probabilistically in cold, dense gas. Gas is determined to be eligible for star formation based on several criteria. The first is the gas density threshold, below which no gas can be converted into a star particle. \texttt{SMUGGLE} adopts a value of 100 cm$^{-3}$, in line with observations of giant molecular clouds \citep[][]{Ferriere2001}. Star formation is also restricted to gravitationally self-bound regions \citep[see][]{hopkins2018fire2}.
Additionally, star formation rates may be computed according to the H$_2$ fraction, though it is usually $\sim 1$ in sufficiently dense gas. 

The probability of an eligible gas cell to be turned into stars is determined via $\dot{M}_\star = $ \esf~$ M_\text{gas}/t_\text{dyn}$, where $t_\text{dyn}$ is the gravitational dynamical time of the gas and $\dot{M}_\star$ its star formation rate. In its default mode, the local SF efficiency parameter \esf~is assigned a value of 0.01 to match the low efficiencies measured in observations \citep{smith2018}, although \citet{hopkins2018fire2} showed that the exact level of feedback-regulated star formation is independent of \esf. We explore in Section \ref{sec:ISM} the effect of \esf~on our \texttt{SMUGGLE} simulations. 

In addition to the default mode described above, \texttt{SMUGGLE} can also be run using the variable efficiency model (\texttt{vareff}), which implements a variable star formation timescale ($t_\text{sfr}$). This quantity, defined as $t_\text{sfr}$ = $M_\text{gas}$ / $\dot{M}_\star$, is varied for each grid cell according to its virial parameter ($\alpha$), which quantifies the cell's ability to resist gravitational collapse via thermal support and gas pressure. The exact parameterization is given by Eqn. \ref{eqn:vareff} below \citep[][]{Padoan2012,Semenov2016}.

\begin{equation}
\label{eqn:vareff}
    t_\text{sfr} = \frac{M_\text{gas}}{\dot{M}_\star}\text{min}\Big(\exp{\big(1.6\sqrt{\alpha/1.35}\big)}, 10^{30}\Big)
\end{equation}

\noindent  This model prioritizes star formation efficiency in highly dense regions. In Section \ref{sec:smuggle_vars}, we investigate both a variable efficiency model, and one that maximizes the local star formation efficiency. Note that since $t_\text{sfr} = t_\text{dyn}/\varepsilon_\text{sf} = M_\text{gas}/\dot{M}_\star$, a parameterization on $t_\text{sfr}$ is equivalent to a parameterization of the efficiency \esf, all other quantities being the same for a given cell.

\subsubsection{Feedback}
Stellar feedback is modeled locally according to several sources including stellar winds, radiation from young stars and supernovae (SNe). Stellar winds due to massive OB and AGB stars contribute to the mass return to the ISM and are taken into account during the pre-processing of the gas. 
Cumulative mass loss from OB stars, as well as energy and momentum returned from both OB and AGB stars are determined via the parameterizations presented in \citet{hopkins2018fire2}, while AGB wind mass transfer is given by \citet{v13feedback}. All the properties determined from the different feedback channels are then injected with corresponding metallicity to the surrounding gas in the rest frame of the star. Stellar winds are a continuous process, and are thus treated continuously across each time step for each star particle. 

Radiative feedback from young stars change the ionization, thermal, and dynamical state of the ISM, pre-processing the media where later SNe will go off. \texttt{SMUGGLE} includes a treatment of photoionization aimed at capturing the formation of HII regions by young, massive stars. Ionizing photon rates from young stellar particles  are calculated by choosing a mass-to-light ratio and average ionizing photon ($> 13.6$\,eV) energy to correspond with a $T = 4\times10^4$\,K blackbody spectrum, consistent with OB type stars. The number of available photons in a given timestep is used to stochastically photoionize neighboring gas cells after accounting for the expected number of recombinations. Photoionized cells are then updated to be fully ionized and placed at a temperature $T=1.7 \times 10^4$\,K. In addition to photoionization, young stars exert radiation pressure on neighboring gas cells, which is calculated according to their optical depth and position within the kernel. Multiple IR scattering is included, by assuming an average opacity $\tau = 10\,Z/Z_\odot$\,cm$^2$\,g$^{-1}$ \citep{hopkins2018fire2}. In the regime of small mass galaxies explored here, photoionization is expected to dominate among the radiation effects on the ISM, lowering the density of gas in the neighborhood of massive stars \citep[][]{Sales2014,hopkins2018fire2}. 

Lastly, we stochastically model the injection of energy and momentum by discrete SN events onto neighbouring gas cells. It is important to note that \texttt{SMUGGLE} resolves individual SN explosions, and as such, the injected rates of energy and momentum are not continuous. The temporal distribution of individual Type Ia events is found by integrating the delay time distribution, which accounts for the approximate lifespan of an 8\,\msun~main sequence star, with rates and energetics consistent with observations \citep[][]{greggio2005} as well as previous implementations in \textsc{Arepo} \citep{v13feedback}, with each event releasing the same mass of ejecta \citep{thielemann2003}. The total number of Type II SNe is found by integrating the Chabrier IMF of each stellar particle. If necessary, we account for PdV work in the (unresolved) Sedov-Taylor phase by applying a momentum boost to match the terminal momentum per SN, which depends primarily on local density and metallicity \citep[e.g. ][]{Cioffi1988}. Energy and momentum are distributed to surrounding gas cells following a kernel weighting and a maximum coupling radius, as described in detail in \citet{Marinacci2019}.

\begin{table*}
    \centering
    \begin{tabular*}{\textwidth}{l @{\extracolsep{\fill}} c c c c c}
        \hline
        
        Name & \rcore~[pc] & $\alpha$ & $M_\star$~[\msun] & Model description \\
        \hline \\ 
        
       \texttt{SMUGGLE} / \texttt{fiducial} & 431.3 & $-0.13$ & 7.76\e{7} M$_\odot$ & default \texttt{SMUGGLE} model \\
        
        SH03            & 160.2 & $-0.52$ & 4.29\e{7} M$_\odot$ & \citet{SpringelHernquist2003} model\\
        
        \texttt{rho0.1} & 324.2 & $-0.05$  & 9.69\e{7} M$_\odot$ & \texttt{SMUGGLE} with reduced gas density threshold, \rhoth~$=0.1$ cm$^{-3}$\\
        
        \texttt{eSF100} & 490.7 & $-0.03$ & 8.39\e{7} M$_\odot$ & \texttt{SMUGGLE} with maximized local SF efficiency, \esf~$= 1$ \\
        
        \texttt{vareff} & 528.3 & $-0.03$ & 9.12\e{7} M$_\odot$ & \texttt{SMUGGLE} with the variable efficiency model, see Sec. \ref{sec:starformation}  \\
        
        \\
        
        \hline
        
    \end{tabular*}
    \caption{List of simulations used in this study. All initial conditions were generated according to \citet{Springel2005} and run for 2 Gyr $h^{-1}$, where we take $h = 0.7$. Our standard resolution initializes a $2.17\times10^{10}$ \msun~ halo with 3\e{7} dark matter particles, and $10^6$ gas particles, corresponding to a baryonic mass per cell of $\sim$850 \msun~and DM mass per cell of $\sim$7200 \msun. We adopt a gravitational force softening of $\epsilon = 16$ pc for all particle types. Also listed are the core radius (measured as described in Section \ref{sec:coresize}), inner DM power law slope $\alpha$, and stellar mass formed (i.e. not including the disk and bulge from initial conditions), all taken at final time.}
    \label{tab:simtable}
\end{table*}

\subsubsection{Variations on the fiducial SMUGGLE model}
\label{sec:methods_variations}

We will explore in Section \ref{sec:ISM} the effect of changing some of the default choices in \texttt{SMUGGLE} and how this affects the formation of dark matter cores and the properties of our simulated dwarfs. The changes will be inspired by results presented previously in the literature, including  \citet{read2016rc,benitezllambay2019,bose2019}. More specifically, we choose to vary the star formation gas density threshold \rhoth~and the local star formation efficiency \esf. 

Table \ref{tab:simtable} summarizes our runs, which include the fiducial \texttt{SMUGGLE} run, SH03, and three variations on \texttt{SMUGGLE} as discussed in Section \ref{sec:smuggle_vars}: (i) \texttt{rho0.1}, using a reduced star formation density threshold of \rhoth~$= 0.1$ cm$^{-3}$;  (ii) \texttt{eSF100}, which maximizes the local star formation efficiency to 100 per cent, \esf $=1$; and (iii) \texttt{vareff}, a variable efficiency model which chooses a value between \esf~$=0.01$ and \esf~$=1$ depending on the density of the surrounding ISM. The fiducial \texttt{SMUGGLE} model implements these parameters with values of \rhoth~$= 100$ cm$^{-3}$ and \esf~$=0.01$.

\subsection{The Springel and Hernquist Model}
In addition to the fiducial \texttt{SMUGGLE} model, we run a simulation with the SH03 model \citep{SpringelHernquist2003}, which forms the basis for the ISM treatments in Illustris \citep{v14illustris,v14nature}, Auriga \citep{Grand2017} simulations, EAGLE \citep{Schaye2015}, APOSTLE \citep[][]{sawala2016}, HorizonAGN \citep[][]{Dubois2014}, SIMBA \citep[][]{Dave2019}, and others. The SH03 model, also run with the {\sc Arepo} gravity and hydrodynamics solver, uses an equation of state treatment of cold gas modelled with a two-fluid approach (cold clouds embedded in a lower density hot gas bath) to describe the interstellar medium. This approach, which has been demonstrated to be successful at modeling the kpc-scale properties of galaxies, has been found to not form dark matter cores \citep[][]{v14dwarfs,bose2019}. 

We explicitly include stellar winds in the SH03 run with the wind velocity calculated directly from the energy and momentum summation of all SN in a given timescale and independent of halo properties. This is different from, for instance, the Illustris or Auriga projects, where the wind velocity is scaled to the dark matter velocity dispersion of the subhalo. Although such scheme is {\it de-facto} closer to the scalings expected for momentum-driven winds \citep{Norman2005} and shown to more accurately reproduce some galaxy and CGM observables \citep[e.g. ][]{Dave2011}, we choose a simpler wind model where no pre-assumptions are made with respect to the properties of the host halo, in an attempt to establish a fairer comparison with the \texttt{SMUGGLE} runs where no input information is required about the galaxy host. Ultimately, the impact of the exact modeling of the winds in our SH03 run is subdominant to the differences imprinted by the modeling of the ISM itself. As is the case in Illustris, Auriga, and other projects mentioned above, the wind particles in the SH03 model are artificially decoupled from the hydrodynamics for a short period of time, while such a treatment is not necessary in our new \texttt{SMUGGLE} prescription where outflows naturally arise from the kinematics and thermodynamics of stellar winds and SN explosions.

\subsection{Isolated Galaxy Setup}
Throughout this paper, we analyze simulations run with different ISM models applied to the same initial conditions (ICs). We initialize an isolated, idealized dwarf galaxy with \mtwo~$= 2.17 \times 10^{10}$\,\msun\ using the method outlined in \citet{Springel2005}. The distribution of dark matter is initialized according to a Hernquist profile \citep[][]{Hernquist1990},
\begin{equation}
\label{eqn:hernquist}
    \rho_\text{dm}(r) = \frac{M_\text{dm}}{2\pi}\frac{a}{r(r+a)^3} \, ,
\end{equation}

\noindent where $a$ is a concentration-dependent scale length. This model is identical to the widely used NFW profile \citep[][]{navarro1996CDMhalos} at small radii ($\rho \propto r^{-1}$), while the power law exponent differs at large radii: $\rho_\text{NFW} \propto r^{-3}$ versus $\rho_\text{Hernquist} \propto r^{-4}$. Both models have been shown to accurately describe the distribution of DM for halos in a cosmological context.

The galaxy itself is initialized with an exponential disk of scale length $h$ for both stars and gas, in addition to a spherical stellar bulge modeled by the Hernquist profile. See Section 2 of \citet{Springel2005} for more details on the model galaxy setup. We choose parameters for our model galaxy consistent with the `Dwarf/SMC' setup described in \citet{Hopkins2011}, which gives a DM dominated dwarf galaxy similar to the pre-infall Small Magellanic Cloud with total baryonic mass $M_\text{bary} = 8.9\times10^8$ \msun, gaseous disk with $M_\text{gas} = 7.5\times10^8$ \msun, and DM halo with $M_{200} = 2\times10^{10}$ \msun~and concentration parameter $c = 15$.

The partitioning of cells in the initial conditions is done by setting the number of gas particles, $N_\text{gas}$, with $N_\text{DM} = 30 N_\text{gas}$, $N_\text{disk} = 0.2 N_\text{gas}$, and $N_\text{bulge} = 0.02 N_\text{gas}$. For the runs analyzed herein, we choose $N_\text{gas} = 10^6$, resulting in a particle mass of $m_\text{bary}\approx 850$ \msun. We choose the same value of gravitational softening for all particle types, with $\epsilon = 16$ pc. We have also run a set of simulations with an order of magnitude lower resolution ($N_\text{gas} = 10^5, \epsilon = 32$ pc) for convergence testing. We find excellent agreement between the two resolution levels tested, as shown in Figure \ref{fig:res_converge}.


\begin{figure*}
    \centering
    \includegraphics[width=\textwidth]{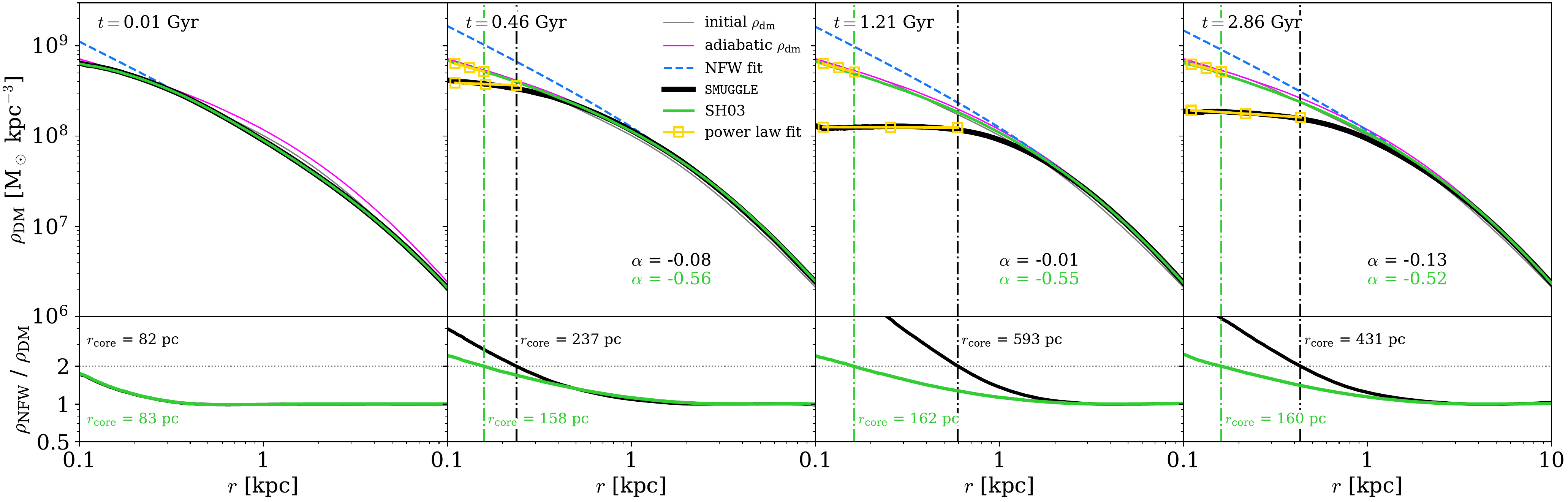}
    \caption{Dark matter density profiles of the isolated dwarf galaxy run with the fiducial \texttt{SMUGGLE} feedback model (black) and the SH03 feedback model (green) at selected times. The top row shows the DM density at each labeled time. Light grey lines represent the DM density profiles at $t=0$, and the blue dashed line is the NFW profile fit to $r$ > 3 kpc to account for variations in the inner region. The core radius \rcore~is defined as the radius where $\rho_\text{NFW}$ / $\rho_\text{DM}$ = 2, and which can be seen in the bottom panels, including the horizontal line at 2. The vertical dashed-dot lines in each panel represented our measured \rcore, which consistently capture the changes in DM density. In addition, power law slopes ($\rho \propto r^\alpha$) are shown in yellow, and are fit for $r_\text{DM}^\text{conv} < r < r_\text{core}$. Values for the convergence radius $r_\text{DM}^\text{conv}$ are typically around 50 pc.}
    \label{fig:rho_panel}
\end{figure*}

\section{Forming Dark Matter cores in SMUGGLE}
\label{sec:smuggle_and_SH03}

We explore the evolution of the dark matter density profile in our simulated dwarf galaxy in Figure \ref{fig:rho_panel}, where each panel corresponds to different times, as labeled. The results of the default \texttt{SMUGGLE} model are shown in the solid black line, which demonstrates a clear flattening in the inner regions corresponding to the formation of a dark matter core in our initially cuspy halo. For reference, we include the initial dark matter distribution in each panel with a solid gray line. 

\begin{figure}
    \centering
    \includegraphics[width=0.47\textwidth]{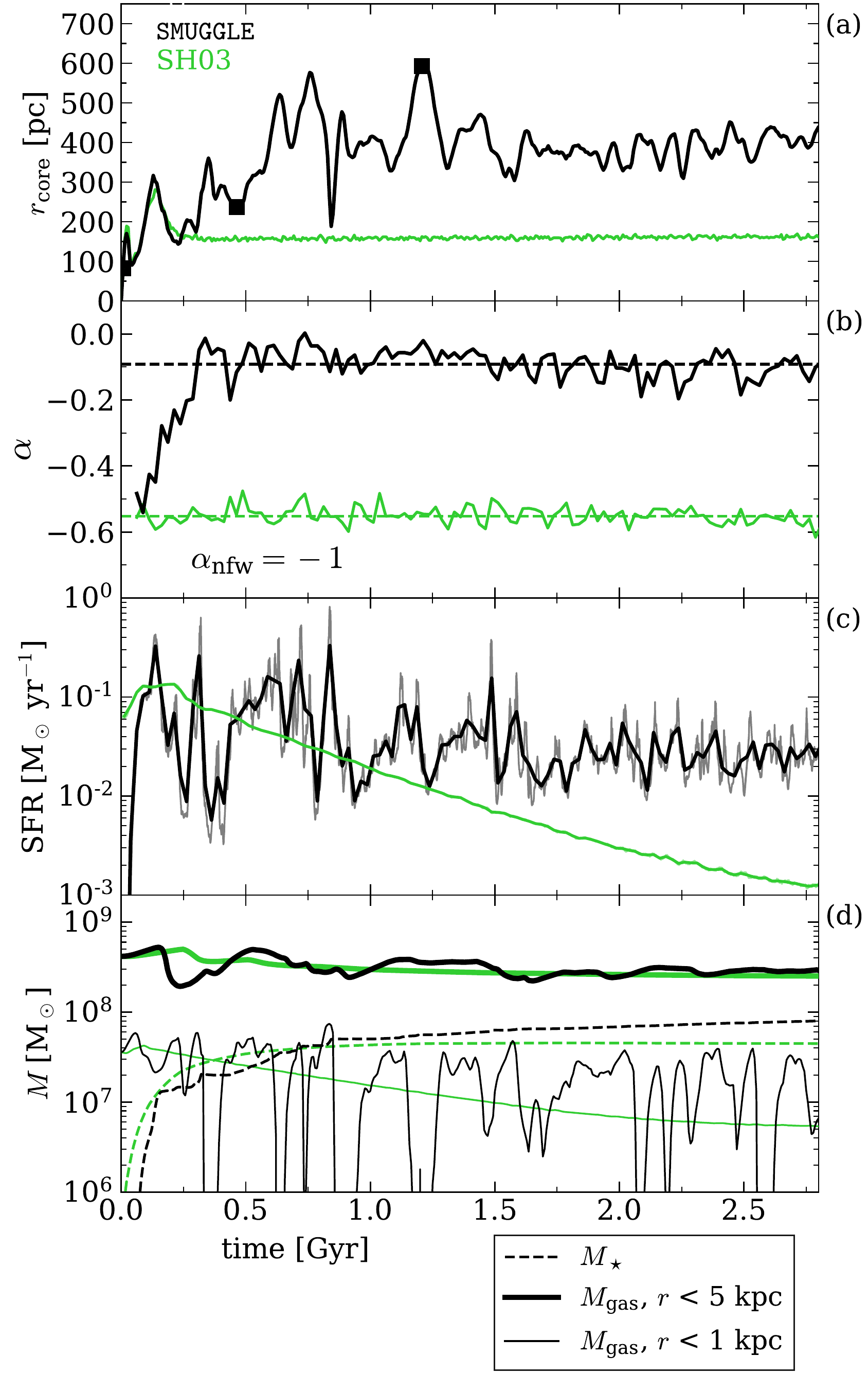}

    \caption{Time-evolving properties of the simulated isolated dwarf galaxy run with the fiducial \texttt{SMUGGLE} model in black and the SH03 feedback model in green. (\textit{a}) Measured core radius \rcore~versus time. Black squares indicate timestamps of density profiles shown in Figure \ref{fig:rho_panel}. See text for definition of \rcore. (\textit{b}) Power law slopes $\alpha$ fitted to $r_\text{DM}^\text{conv} < r < r_\text{core}$, binned with $\Delta t$ = 25 Myr. Dashed lines indicate the average slope for $t > $ 0.5 Gyr to account for initial relaxation effects. The \texttt{SMUGGLE} model results in a very flat inner profile ($\alpha\sim-0.1$) which extends over a larger portion of the galaxy with \rcore $\sim$ 400 pc, in contrast to the steeper ($\alpha\sim-0.6$), more concentrated (\rcore $\sim$ 150 pc) profile formed by SH03.  (\textit{c}) Star formation rate (SFR) versus time. The SFR is smoothed over $\Delta t = $ 25 Myr bins. We find that \texttt{fiducial} produces a substantially burstier star formation history (SFH) than SH03, and that the average magnitude of SFR for SH03 agrees with that of \texttt{fiducial} in early times, but declines to much smaller levels after $t\approx$ 1.5 Gyr. (\textit{d}) Stellar mass (\mstr, dashed), gas mass within $r < 5$ kpc (solid, thick), and gas mass within $r < 1$ kpc (solid, thin). \texttt{SMUGGLE} results in frequent and significant changes in gas mass in the inner regions, while the gas mass < 1 kpc in SH03 smoothly decreases.}
    \label{fig:rcore_sfr}
\end{figure}

\subsection{A consistent method for core size measurements}
\label{sec:coresize}

\subsubsection{Caveats \& numerical effects}

Figure \ref{fig:rho_panel} shows density profiles for various runs implementing the same ICs. We find the best fit NFW profile to the outer ($r > r_{\rm fit} = 3$\,kpc) dark matter distribution. The bottom panels in Figure \ref{fig:rho_panel} show the ratio between the analytic NFW fit and the measured DM density in the fiducial \texttt{SMUGGLE} simulations (solid black lines). Although in the outskirts the simulated profiles are very well described by the NFW fits ($\rho_{\rm NFW}/\rho_{\rm dm} \sim 1$), in the inner regions the analytic profile clearly overestimates the dark matter density in all cases. This is partially due to adiabatic contraction, demonstrated by the magenta line. In the case of SH03, feedback is not capable of producing further changes in the DM distribution, resulting in a profile almost identical to the adiabatic run, while the \texttt{SMUGGLE} model is able to produce an extended region of constant density by later times. Additionally, the shape of the galaxy can affect the resultant DM distribution. In the case of disks, this can lead to shallower central profiles \citep[][]{Burger2021}. 

We note that numerical effects can spuriously transfer kinetic energy between particles of different masses, such as our gas and DM particles \citep[][]{Ludlow2019a}. A thorough investigation of the effects of gravitational softening and `numerical feedback' have been presented in \citet{Ludlow2019b,Ludlow2020}. While we adopt softening on the order suggested by \citet{vandenBosch2018} -- approximately three times lower than the convergence radius $r_{\rm dm}^{\rm conv}$ -- it is possible that spurious energy transfer between DM and baryonic particles via 2-body interactions contributes to the observed halo expansion. However, our tests are designed to isolate the effects of feedback. Numerical effects will be present in all our simulations, including the adiabatic and SH03 runs, but the methods of feedback coupling to the ISM vary. As such, our claims are about the differential effects between feedback implementations, not predictions of the absolute core sizes expected within dwarf galaxies in a cosmological context.

\subsubsection{Core size measurement}
Following \citet{benitezllambay2019}, we define \rcore~as the location where the simulated dark matter density is a factor of 2 lower than the extrapolated best-fit NFW profile, $\rho_{\rm NFW}/\rho_{\rm dm} = 2$. However, we note that the authors compare against a low-threshold run rather than an NFW. Hydrodynamic relaxation may lead to a difference in predicted core radius. The measured \rcore~is indicated with a vertical dashed line and listed in the lower panels. 

This definition is robust to variations on $r_{\rm fit}$ in the range $1 - 10$ kpc (see Figure \ref{fig:rcore_slope_compare_nfwdcut} in the Appendix). Figure \ref{fig:rho_panel} shows that the density profile within \rcore~for the fiducial \texttt{SMUGGLE} run is nearly flat at later times. We quantify this by finding the slope $\alpha$ of a power-law fit to the dark matter density between the convergence radius $r_{\rm dm}^{\rm conv}$ and \rcore, where $r_{\rm dm}^{\rm conv}$ is defined as the radius containing 200 DM particles \citep[as in ][]{Klypin2001,hopkins2018fire2}, and is typically around 50 pc in size. For reference, the measured slopes $\alpha$ are quoted in each panel. 

While the initial DM distributions of our simulations follow a Hernquist profile, we find no difference in measured core radius when using Hernquist or NFW parameterizations, consistent with the intended similarity between the fits for $r \ll r_{200}$. While some choices of our methodology are arbitrary, we find that it consistently produces an accurate characterization of the physical extent and slope of the constant density inner regions. We show in the Appendix that core formation is well converged and robust to numerical choices, such as resolution and $r_{\rm fit}$ (see Figs.~\ref{fig:res_converge} and \ref{fig:rcore_slope_compare_nfwdcut}).


\subsection{Halo response to SMUGGLE versus SH03 models}
Interestingly, and in contrast to previous results of model implementations within {\sc Arepo} \citep[e.g.][]{Marinacci2014,Chua2019,bose2019}, we find that the new \texttt{SMUGGLE} model develops a well-defined constant-density core with radius $200 - 600$\,pc in our idealized $M_{200} \sim 10^{10}$ \msun~dwarf halo. In comparison, the same initial setup run with the SH03 model does not robustly form a core. 

In practice, our method suggests \rcore~$\approx 175$\,pc (see bottom panels) for the SH03 run, although this is more consistent with a relaxation effect than a true dark matter core achieved by repeated perturbation of the potential. This is further supported by the inner slope $\alpha$, which is far from being a flat constant density distribution ($\alpha \sim 0$) as found for our fiducial \texttt{SMUGGLE} run and instead remains steep ($\alpha \sim -0.55$), consistent with that of the initial condition over a similar distance range. In addition, we have run an adiabatic (i.e. no star formation or feedback) version of the same initial setup for $t \sim 0.7$\,Gyr. By our methods, we calculate time-averaged values of \rcore~$= 150$\,pc and $\alpha = -0.57$ for the adiabatic run, indicating that the behavior seen in SH03 is consistent with relaxation and is not representative of a feedback-induced core. Note the similarity between the green SH03 and magenta adiabatic curves in Figure \ref{fig:rho_panel}.

We therefore find that the SH03 ISM treatment does not create a core, in agreement with previous studies implementing similar models \citep[e.g.][]{Marinacci2014,bose2019} while the new ISM treatment \texttt{SMUGGLE} results in clear halo expansion. The measured core extends over several hundred pc, which is well beyond the gravitational softening for the dark matter $\epsilon=16$\,pc or the convergence radius $r_{\rm DM}^{\rm conv} \approx 50$\,pc.

A more detailed description of the time evolution for the core is shown in the panels (a) and (b) of Figure \ref{fig:rcore_sfr}, showing the core radius \rcore~and the power law slope $\alpha$ of the inner region $r^{\rm DM}_{\rm conv} < r < r_\text{core}$ of the dark matter density profile. In \texttt{SMUGGLE}, the core radius grows during the first Gyr, after which it settles on an average \rcore~$\sim 400$\,pc with fluctuations. The slope flattens from $\alpha = -0.55$ to $-0.09$ in the first half Gyr, where it remains for the rest of the simulation. In contrast, SH03 relaxes into a stable density distribution with  \rcore $\sim 160$\,pc and no significant change in slope, resulting in a cusp rather than a core.


Panel (c) of Figure \ref{fig:rcore_sfr} compares the star formation histories in the \texttt{SMUGGLE} and SH03 runs. The rapid fluctuations in the \texttt{SMUGGLE} run are sustained throughout the $\sim 3$\,Gyr of run time, though with decreased burstiness after $t \sim 1$\,Gyr. This contrasts the smoother SFR from the SH03 ISM model. In fact, SH03 shows a declining SFR, likely due to the lack of cold inflows and depletion of all eligible star forming gas. The cooling implementation of SH03 results in an effective temperature floor of $\sim 10^4$\,K, such that, with the lack of cold inflows, no new gas is able to condense to sufficiently high densities to fuel star formation. As a result, the final stellar mass formed in \texttt{SMUGGLE} is $\sim 50\%$ larger compared to SH03.

Note that this burstiness in the star formation of \texttt{SMUGGLE} is associated to fluctuations on the gas mass in the inner 1 kpc (Figure \ref{fig:rcore_sfr}, panel d), while SH03 simply depletes the gas content in this region. As discussed in \citet{pontzengovernato2012}, such mass fluctuations in short timescales can cause the local gravitational potential to non-adiabatically change resulting in the expansion of dark matter orbits and, consequently, on the formation of a lower density core. 
In the case of \texttt{SMUGGLE}, although the gas content is changing very abruptly in the very inner regions (thin) and less so outwards, the mass fluctuation can be discerned quite far out into the main body of the galaxy, $r \sim 5$\,kpc.

What is driving these differences between the ISM models? Discussions in the literature have cited rapid fluctuations of the potential in the inner regions \citep{navarro1996,pontzengovernato2012}, burstiness of star formation rates \citep{madau2014,chan2015,tollet2016,Dutton2019}, and high gas densities such that it dominates the central potential \citep{benitezllambay2019}. These features are all present in the \texttt{SMUGGLE} treatment but not in the SH03-like models, explaining why core formation is achieved in \texttt{SMUGGLE} but not in previous ISM treatments in {\sc Arepo}. 

\begin{figure}
    \centering
    \includegraphics[width=0.45\textwidth]{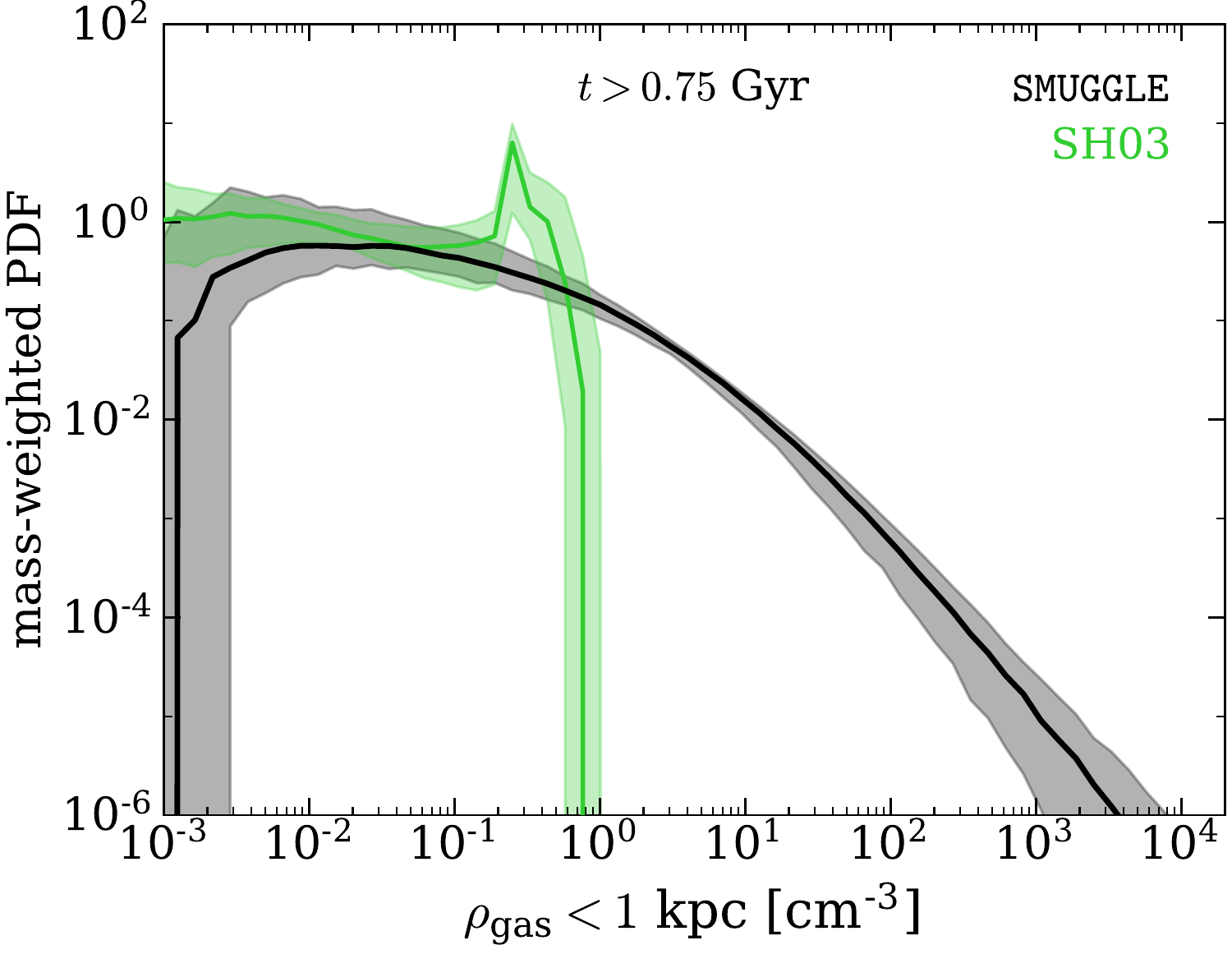}
    \caption{Median mass-weighted probability density function of gas density for $t > 0.75$ Gyr for the inner 1 kpc, with shaded regions representing the 68 per cent confidence interval in each $\rho$ bin. The fiducial \texttt{SMUGGLE} run is able to achieve gas densities of $>10^3$ cm$^{-3}$, while SH03 is unable to obtain densities $> 1$ cm$^{-3}$. The higher densities achieved by \texttt{SMUGGLE} allow its gas to gravitationally influence the DM to a stronger degree than in SH03. }
    \label{fig:rhogas1kpc_fidill}
\end{figure}

Figure \ref{fig:rhogas1kpc_fidill} shows the time-averaged gas density distribution within 1 kpc for each run. This distribution is calculated with equal logarithmically spaced bins between $\rho_\text{gas}=10^{-6}\,\text{cm}^{-3}$ and $\rho_\text{gas}=10^{6}\,\text{cm}^{-3}$ at each snapshot. The median gas density is then calculated for each bin to construct the final gas density distribution, with standard deviation about the median shown as shaded regions. 

As a result of the molecular cooling and other physics modeled in \texttt{SMUGGLE}, the typical gas densities achieved in \texttt{SMUGGLE} can be several orders of magnitude higher than in SH03. This run results in very few gas particles denser than $\rho = 1$\,cm$^{-3}$ (green curve) while about half of the gas in the \texttt{SMUGGLE} run is above that threshold and up to $\sim 10^4$\,cm$^{-3}$. The high gas densities achieved by \texttt{SMUGGLE} are instrumental in gravitationally perturbing the dark matter to create cores, while the wide range of densities reached in the inner 1\,kpc indicates repeated disruption of dense gas from feedback in central star forming regions, maintaining a multi-phase nature that compares well with observations of real galaxies. While models based on an equation of state ISM treatment might be able to reproduce and predict statistical properties of galaxy populations as well as large-scale structure with remarkable success \citep[e.g. ][]{v14illustris, Marinacci2014, Schaye2015, sawala2016, Grand2017, Pillepich2018}, they cannot capture the interplay between DM and baryons on small scales, where the contribution of baryons to the gravitational potential is significant.


\section{The Effect of the ISM Model Parameters}
\label{sec:ISM}

\subsection{Variations on \texttt{SMUGGLE}}
\label{sec:smuggle_vars}

In addition to the fiducial \texttt{SMUGGLE} model and SH03, we have run three simulations using the same initial conditions with variations on key parameters in the \texttt{SMUGGLE} ISM model: (i) \texttt{rho0.1} reduces the star formation gas density threshold\footnote{The H$_2$ star formation requirement discussed in Section \ref{sec:SMUGGLEmodel} was lifted to allow the density threshold to take full effect.} from the fiducial value of \rhoth~= 100 cm$^{-3}$  to \rhoth~= 0.1 cm$^{-3}$ to mimic the value used in simulations such as SH03 and EAGLE \citep[][respectively]{v13feedback,crain2015}; (ii) \texttt{eSF100} increases the star formation efficiency from the fiducial value of \esf~= 0.01 to \esf~= 1 to compare with FIRE \citep{hopkins2018fire2}; and (iii) \texttt{vareff}, which parameterizes \esf~(see Section \ref{sec:methods_variations}, Eqn \ref{eqn:vareff}) to maximize star formation in dense, self-gravitating gas clouds. Table \ref{tab:simtable} summarizes these runs and their key features.


\begin{figure}
    \centering
    \includegraphics[width=0.45\textwidth]{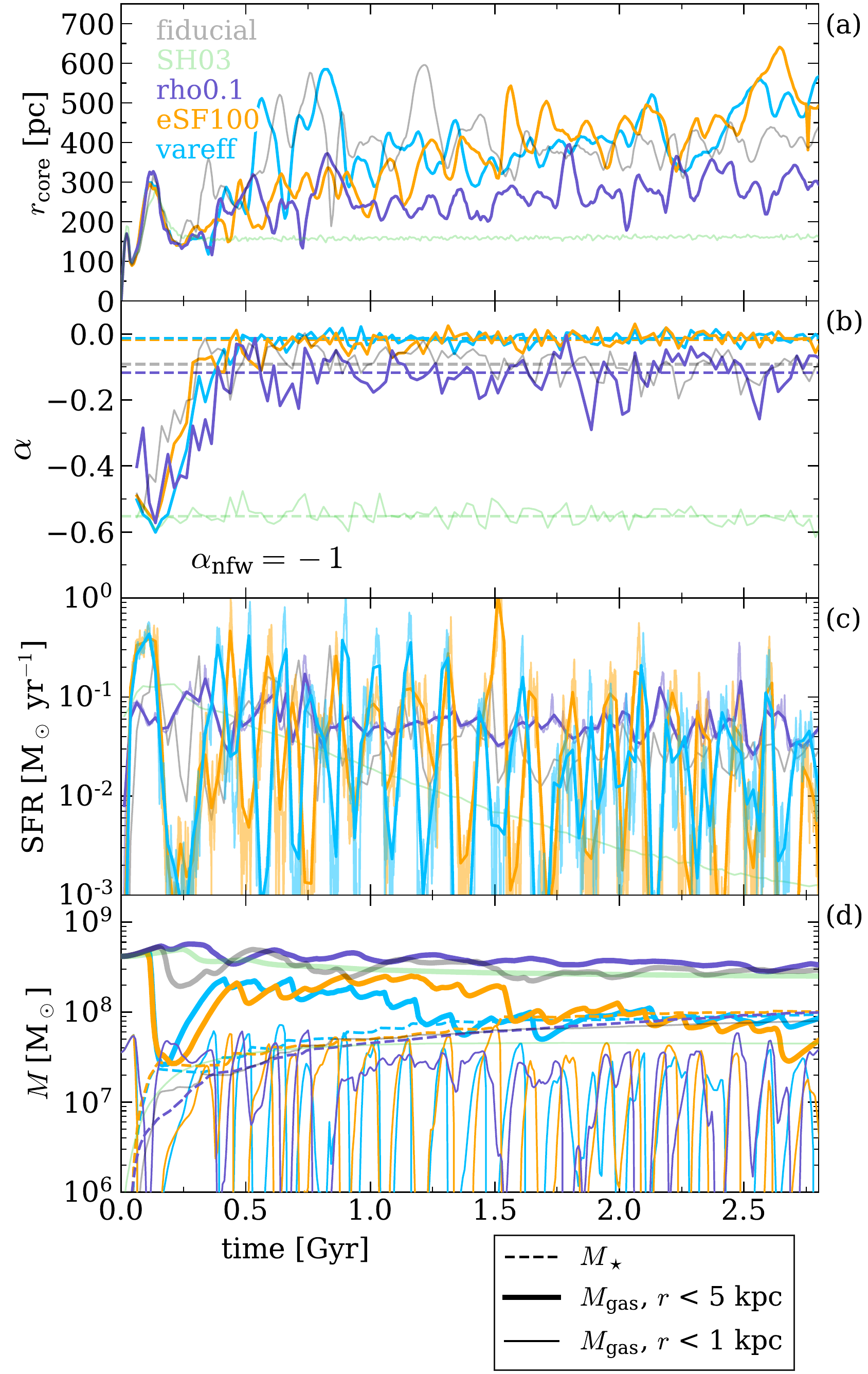}
    \caption{Selected properties for \texttt{rho0.1} (purple), \texttt{eSF100} (orange), and \texttt{vareff} (blue), as in Figure \ref{fig:rcore_sfr}, including faint lines for \texttt{fiducial} and SH03. All variations on the \texttt{SMUGGLE} model are able to form flattened DM cores between approximately 250--400\,pc in extent and with $\alpha\sim -0.1$--$0$. \texttt{rho0.1} shows the least bursty SFR of the \texttt{SMUGGLE} runs, while both \texttt{eSF100} and \texttt{vareff} have SFRs that are significantly burstier than the fiducial \texttt{SMUGGLE} model. Remarkably, all \texttt{SMUGGLE} runs converge in \mstr~within $\sim$20\%, despite differences in SFR and gas content. The effect of different SFRs can be seen in the bottom panel as sharp jumps in \mstr~and decreases in $M_\text{gas}$ (outflows), or the lack thereof. We see that the high efficiency runs undergo repeated outflows, slowly depleting their gas reserves, while \texttt{fiducial}, \texttt{rho0.1}, and SH03 retain a majority of their original gas content.}
    \label{fig:rcore_varISM}
\end{figure}

Figure \ref{fig:rcore_varISM} shows time-dependent properties of the variations on \texttt{SMUGGLE}, with the original two runs shown in faded, thin lines. The core radius and slope are shown in Panels (a) and (b). We find that all \texttt{SMUGGLE} runs form clearly defined cores, with shallow slopes and core sizes larger than demonstrated in SH03. We find that time-averaged ($t>0.75$ Gyr) values of \rcore~vary from $275 - 400$ pc in extent, with slopes of $\alpha=-0.07\pm0.06$. This is within the range of core sizes observed for dwarf galaxies in the literature, with typical values of $\alpha = -0.2\pm0.2$ \citep{deblok2001,Oh2011,Oh2015}. 

We find variation between the different \texttt{SMUGGLE} runs. The low threshold \texttt{rho0.1} forms the smallest \rcore, as expected, though much more of a robust core than the mild expansion seen in SH03. Interestingly, the high efficiency run \texttt{eSF100} appears to form its core slower than \texttt{fiducial}, but ends up with a larger core by the final time. The variable efficiency run \texttt{vareff} forms its core early on -- similar to \texttt{fiducial} -- but continues to grow at later times. These variations, however, are relatively minor. The primary distinction between the fiducial \texttt{SMUGGLE} model and its two increased efficiency variations appears to the continued growth of the core over time as a result of the sustained burstiness of their star formation. This is likely due to the increased energy injection into the ISM via the efficient star formation and SN feedback. That is, a much higher fraction of gas that is eligible to turn into star particles is converted. For contrast, the fiducial \texttt{SMUGGLE} model only turns $\sim$ 1 per cent of the eligible gas into stars (on an average, not per-particle basis), in accordance with observations of GMCs \citep[][]{smith2018}. These strong blow-outs represent a somewhat different, more violent mode of core formation than exhibited in the fiducial run, which experiences smaller, more frequent outbursts. Convergence among runs to universally shallow slopes is notable. However, we do still observe that the higher efficiency runs \texttt{eSF100} and \texttt{vareff} form slightly shallower cores with $\alpha\sim -0.03$, while \texttt{rho0.1} and \texttt{fiducial} form cores with $\alpha\sim -0.1$. 

Panels (c) and (d) of Figure \ref{fig:rcore_varISM} show the SFR, stellar mass, and gas mass versus time for all runs. The SFRs we observe in the new \texttt{SMUGGLE} models are within expectation. The \texttt{rho0.1} run maintains a higher average SFR due to a lower \rhoth, which effectively  increases the amount of gas that is eligible for SF at any given timestep. Meanwhile, the higher efficiency runs see extremely bursty star formation histories due to a cycle of intense star formation, feedback that blows gas out of the inner regions, and re-accretion of gas to eligible SF densities. Despite these differences in star formation, we find excellent convergence in \mstr~for all \texttt{SMUGGLE} runs, with all runs reaching a final value within $\sim$20\% of one another. 

However, we do find differences in gas content and nature of outflows between these runs. We see that \texttt{rho0.1} retains more of its gas within 5 kpc than \texttt{fiducial} while also undergoing fewer and shallower outflows (seen as dips in the gas mass). In stark contrast, the highly efficient runs lose a majority of their initial gas content by the end of the simulation, undergoing frequent and larger outflows than either \texttt{rho0.1} or \texttt{fiducial}, retaining only $\sim$20\% of their original gas mass by $t=2.0$ Gyr$h^{-1}$.

\begin{figure}
    \centering
    \includegraphics[width=0.45\textwidth]{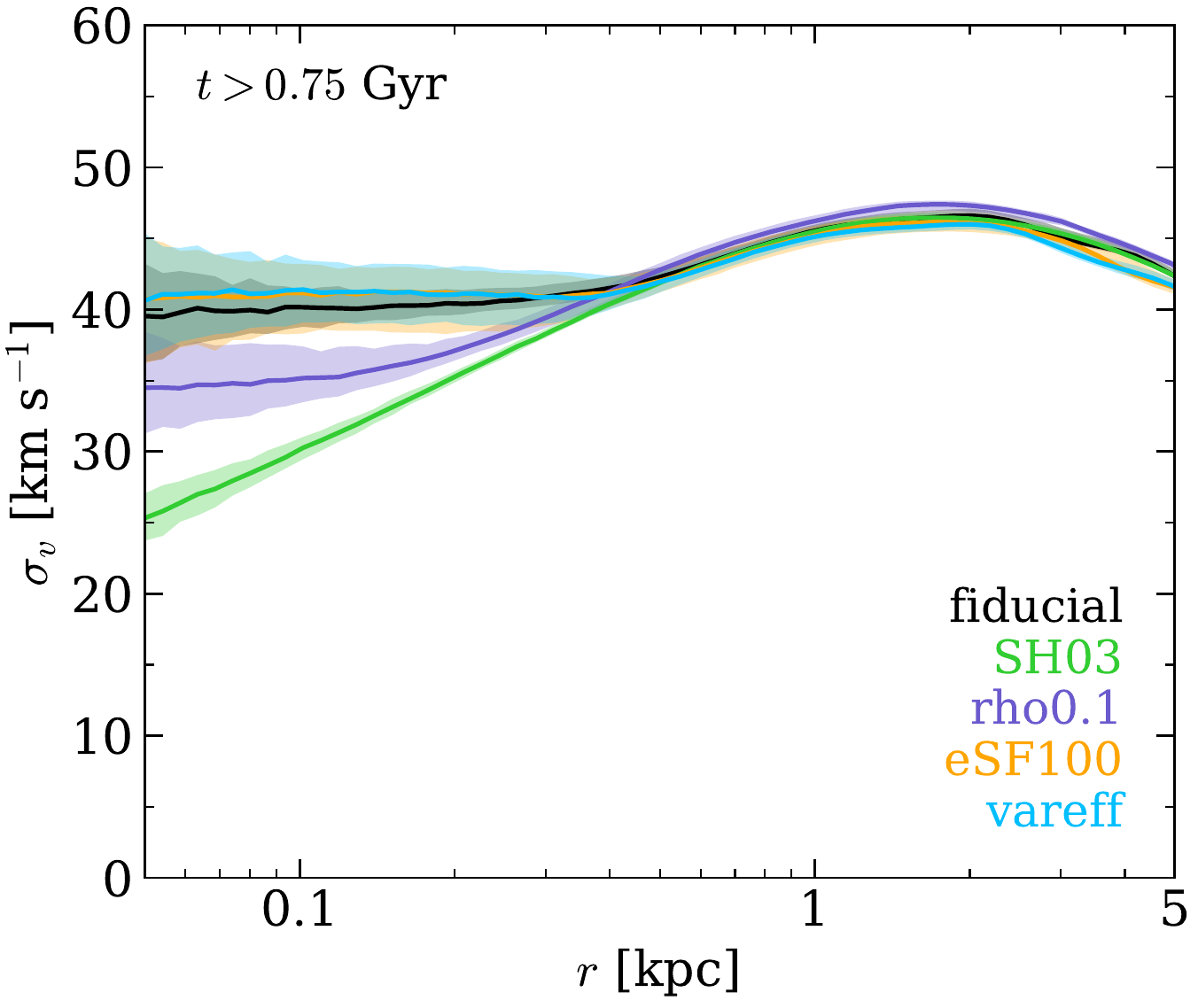}
    \caption{Time-averaged dark matter velocity dispersion profiles for each run. We find that the high efficiency variations on \texttt{SMUGGLE} approximately reproduce an isothermal (constant $\sigma_v$) core in the inner regions, while the SH03 run produces a decreasing profile similar to an NFW.}
    \label{fig:sigma_dm}
\end{figure}

Figure \ref{fig:sigma_dm} shows the DM velocity dispersion for all runs, averaged over the final 0.5 Gyr of the simulations. We find results roughly as expected: the velocity dispersion of SH03 is consistent with a cuspy NFW profile, while the \texttt{SMUGGLE} runs form ever-flatter inner profiles, approaching the constant-$\sigma$ signature of an isothermal profile with the higher efficiency runs, as expected from self-interacting dark matter models \citep[][]{Vogelsberger2012,Rocha2013,TulinYu2018,Burger2019}. 
While it is interesting to see isothermal velocity dispersion profiles generated as a result of baryonic feedback, these results are not identical with expectation from SIDM. For example, profiles in SIDM are isothermal to much larger radii, then immediately decline, whereas the contribution from baryons results in a sizable bump at intermediate radii with a smoother tail. This may a possible avenue to distinguish SIDM from baryonic feedback \citep[][]{Fitts2019}. Additionally, the isothermal profiles seen in the \texttt{SMUGGLE} runs demonstrate that they are not in dynamical equilibrium, an effect we discuss in Section \ref{sec:diversity}.

\begin{figure}
    \centering
    \includegraphics[width=0.45\textwidth]{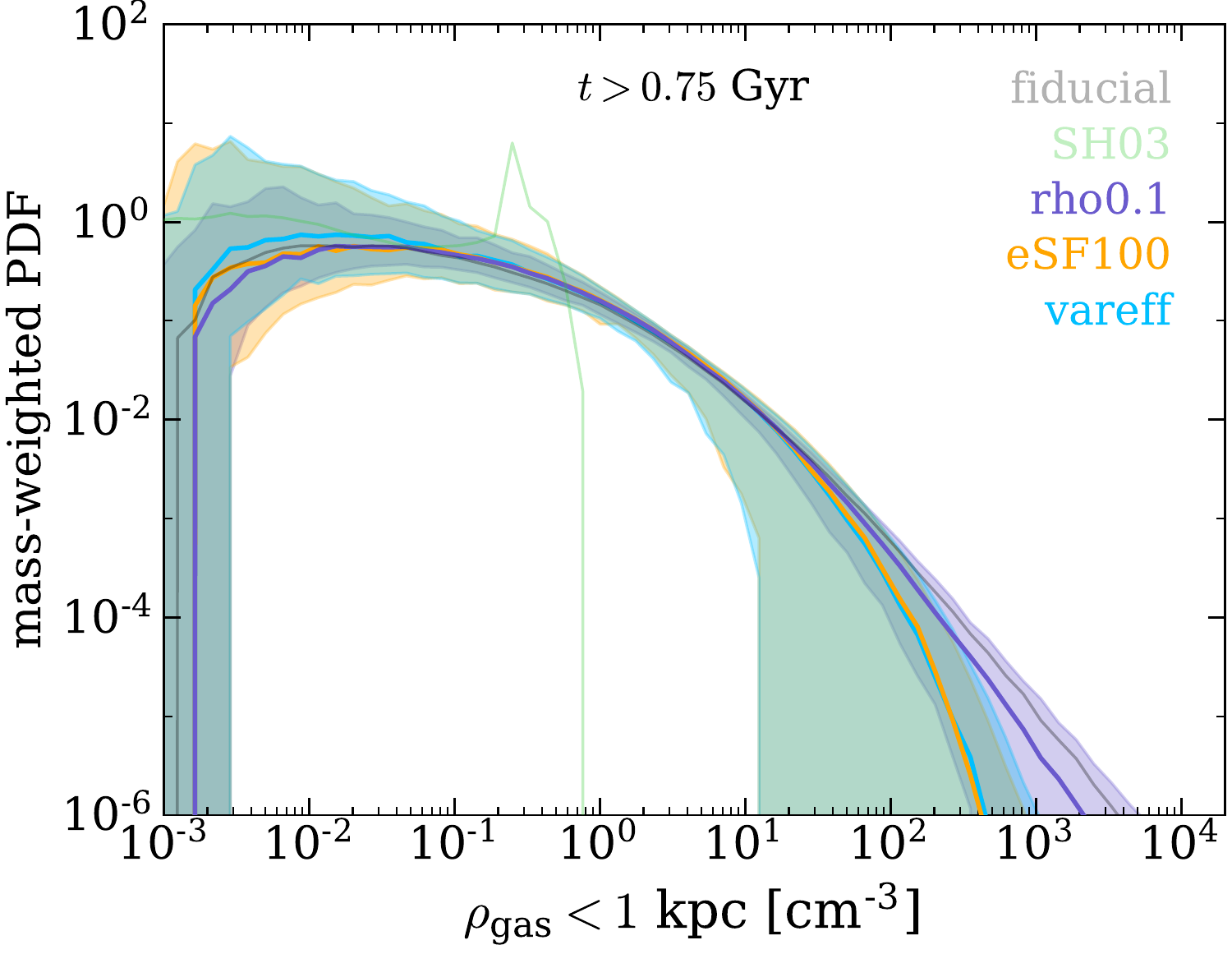}
    \caption{Median gas density distribution for each run over the run time of the simulation after $t=0.75$ Gyr, with shaded regions representing the 68 per cent confidence interval in each $\rho$ bin. Both \texttt{fiducial} and \texttt{rho0.1} are able to produce an ISM with a substantial fraction of the gas above their star formation thresholds, while the median gas densities achieved by \texttt{eSF100} and \texttt{vareff} demonstrate a more rapidly decreasing high density tail. This is a result of different star formation efficiencies: in the high efficiency runs, gas that reaches \rhoth~is quickly turned into stars, while low efficiency preserves a component of highly dense gas.}
    \label{fig:rhohist1kpc_varISM}
\end{figure}

As discussed previously, examining average gas densities can be a useful exercise to understand the behavior of both the DM and the baryons. Figure \ref{fig:rhohist1kpc_varISM} shows the same time-averaged gas density calculation as Figure \ref{fig:rhogas1kpc_fidill}, but for all runs, including shaded regions for standard deviations. 
Interestingly, we find that \texttt{rho0.1} is able to produce gas densities well above its star formation threshold of \rhoth$=0.1$ cm$^{-3}$, with an almost identical distribution to \texttt{fiducial}, though slightly favoring lower densities. In contrast, the runs with higher efficiencies (\texttt{eSF100} and \texttt{vareff}) are limited to gas densities at or near the standard value of \rhoth~= 100 cm$^{-3}$, with slightly lower values in the fully efficient \texttt{eSF100} than in the selectively efficient \texttt{vareff}. 
The changes in the high-density tail between fiducial \texttt{SMUGGLE} model and \texttt{eSF100} are consistent with results from \citet{Li2020}, who investigated the effects of this parameter on GMCs in MW-mass galaxies.



\subsection{The role of modeling parameters}
As discussed in Section \ref{sec:smuggle_and_SH03}, we find that the same isolated galaxy setup run with the SH03 feedback model does not form cores due to its relatively low density gas and its lack of bursty star formation. It is generally claimed that these features are governed by the choice of \rhoth~in the model \citep{pontzengovernato2012,bose2019,benitezllambay2019}, however, the clear differences between SH03 and \texttt{rho0.1}, both of which implement a low density threshold of \rhoth~= 0.1 cm$^{-3}$, demonstrate that the physics of core formation is dependent on factors beyond this parameter.

The physical differences between these runs is clear: \texttt{rho0.1} has somewhat bursty star formation, dense gas, and SN-driven outflows of gas from the central regions, while SH03 has monotonically decreasing SFR, sparse gas, and no discernible feedback-driven outflows. If both runs implement \rhoth~= 0.1 cm$^{-3}$ yet achieve such different outcomes, other differences in subgrid physics must be to blame. The unstructured ISM of the SH03 feedback model is a result of its conception as a model for large-scale structure simulations, and is not particularly well suited for studying small-scale structures of galaxies and their halos, such as DM cores. The detailed ISM model implemented in \texttt{SMUGGLE} is able to achieve much higher gas densities, resolving multiple physical gas phases at smaller scales, as well as achieving the bursty star formation necessary to form cores. 

The difference in density achieved by these two runs (Figure \ref{fig:rhohist1kpc_varISM}) therefore points to two facts: (1) the physical gas density achieved by a simulation is not solely governed by \rhoth, especially when using local star formation efficiencies lower than 100 per cent and (2) gas density and star formation burstiness (which drive outflows and subsequently core formation) are a product of the subgrid physics model as a holistic enterprise, including processes such as cooling physics and self-shielding, as well as resolution to the extent that such processes are resolution-dependent, rather than any individual parameter. However, changes in relevant parameters, as demonstrated here and in many other works, \citep[e.g.][, Burger et al. \textit{in prep.}]{pontzengovernato2012,benitezllambay2019} do indeed produce observable differences within the same overall modeling scheme.



In their seminal work, \citet{pontzengovernato2012} compare cosmological zoom simulations run with the SPH \textsc{gasoline} code \citep{Wadsley2004,Stinson2006} run with two different value of \rhoth, corresponding to our fiducial value of \rhoth = 100 cm$^{-3}$ and a low threshold run with \rhoth = 0.1 cm$^{-3}$, as in our \texttt{rho0.1} run. They find that the low threshold run does not form a core, yet the high threshold run does, comparing the same overall ISM model in both cases. They point out that fluctuations in potential result in the expansion of the orbits of DM particles in the inner halo. We emphasize in this discussion that it is the ability of a model to create these physical density fluctuations that matters in producing DM cores. 

As noted by \citet{benitezllambay2019}, it is indeed surprising that few systematic tests of the star formation density threshold have been conducted by this time. The authors investigate the effect of a variety of values for \rhoth~spanning 0.1 cm$^{-3}$ up to 640 cm$^{-3}$ for cosmological halos in the EAGLE simulations \citep{crain2015}. They find that core formation is maximized for values between 1 cm$^{-3}$ and 160 cm$^{-3}$, but find smaller cores for smaller values of \rhoth~due to the lack of gravitationally dominant gas, and also for larger values due to the inefficiency of EAGLE's feedback model in this regime. They identify that core formation in dwarf galaxies is not sufficiently explained by either burstiness of star formation or strong outflows of gas within the EAGLE model. Instead, they point to features in the SFH of different halos that produce differences in outcomes of core sizes.

A similar investigation, though over a smaller range of threshold values, was conducted by \citet{Dutton2019,Dutton2020} for the NIHAO simulation project \citep{wang2015nihao}. They find that, of their halos run with \rhoth = 0.1 cm$^{-3}$, 1 cm$^{-3}$, and 10 cm$^{-3}$, only those with \rhoth = 10 cm$^{-3}$ and stellar mass to halo mass ratio of 0.1--1\% underwent strong expansion, in agreement with the trend pointed out in \citet{dicintio2014SMHM}. Further, they identify that variability in star formation feedback must occur at sub-dynamical time-scales to produce expansion of the DM halo.

In the case of \textsc{gasoline}, a change in density threshold was able to predictably alter the outcome of core formation. The picture is somewhat more complex for EAGLE and NIHAO, which find that core formation, while increasing with \rhoth, is further dependent on SFH, timescale of burstiness, and halo mass, among other things. All these studies examined cosmological simulations. Our idealized numerical experiments seek to eliminate the complexities of cosmological runs, which produce substantial halo-to-halo variations in \mstr/\mtwo, SFH, merger histories, gas fractions, etc. These are all important factors in understanding the diversity of observed galaxies, but can serve to obscure the impact of modeling choices.

Our idealized \texttt{SMUGGLE} runs produced cores for both the fiducial threshold of \rhoth~= 100 cm$^{-3}$ and the lowered threshold of \rhoth~= 0.1 cm$^{-3}$, though \texttt{rho0.1} did produce a somewhat smaller core radius ($\sim 300$ pc, versus $\sim 400$ pc for the fiducial run). When compared to the cuspy profiles of SH03, the core size within these two variations of \texttt{SMUGGLE} can be considered quite similar. This similarity in core size and shape between the two \texttt{SMUGGLE} variations makes sense in light of their achieved physical gas density distributions (Figure \ref{fig:rhohist1kpc_varISM}) versus the highly truncated distribution of SH03, which is incapable of producing $\rho_\text{gas} \gtrsim 1 $cm$^{-3}$. With an initial mean DM density of $\sim$ 4 $m_{\rm p}$cm$^{-3}$ within 1 kpc, it is clear that, even if SH03 produced fluctuations in gas mass within this region, it would be insufficient to perturb the DM potential. 

Another factor that impacts the physics of core formation is the ability of the gravity solver to resolve the free-fall timescale of gas in the centermost star-forming regions of the galaxy. When larger softening lengths are used, the collapse of gas into dense clouds is delayed, and the resulting star formation process will be smoothed out. This leads to fewer discrete star formation events, and a reduction in both the burstiness of star formation and maximum gas density achieved in star forming regions, limiting the growth of cores.

\begin{figure*}
    \centering
    \includegraphics[width=\textwidth]{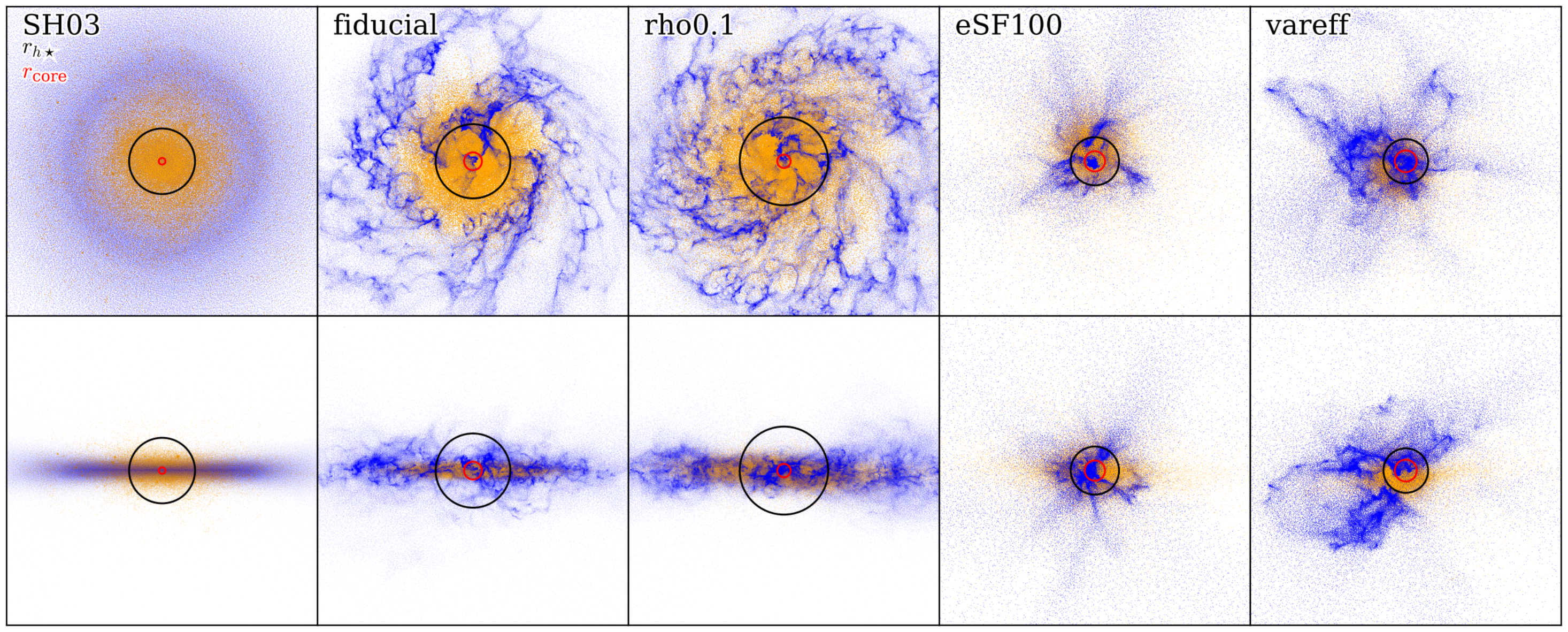}
    \caption{Face-on (top row) and edge-on (bottom row) projections of stars (orange) and gas (blue) for all runs. We only include star particles that were formed after the simulation began, not the old disk and bulge components from the initial condition. Each panel edge is 15 kpc in length, with the half stellar mass radius \rh shown in black (again, only new stars), and the core radius \rcore~shown in red. Both the fiducial run and \texttt{rho0.1} maintain fairly well behaved disks, though with somewhat more disturbance and fragmentation in \texttt{fiducial} as well as a more compact distribution of stars, while \texttt{rho0.1} has a more extended stellar distribution with a smaller core radius. Both \texttt{eSF100} and \texttt{vareff} show a highly disturbed ISM, with gas extending much further in the $z$-direction (perpendicular to the disk). Both galaxies have more compact stellar distributions than the fiducial run.}
    \label{fig:10panelprojection}
\end{figure*}

We emphasize that it is the ability of a model to produce both sufficiently dense gas and sufficient variation in baryonic mass in the inner regions of a halo that will allow it to form cores. The ability of \rhoth~to affect these physical phenomena depends (i) on how the chosen modeling prescriptions modulate the effect of that parameter on star formation, (ii) on how energy injection and dissipation distribute energy throughout the ISM, and (iii) on the interplay between resolution and all of the above. In short, the precise role of \rhoth~in core formation is model-dependent. For example, \texttt{SMUGGLE} produces similar inner gas density distributions regardless of the adopted value of \rhoth, and forms a feedback-induced core in all our explored variations. While the density threshold parameter is a commonly used parameter in ISM models, making it an appealing avenue for study, more attention must be given to the differences between modeling prescriptions with respect to their resulting physical properties (such as the gas density distribution and fluctuations in baryonic mass) before the effects of individual parameters can be understood in proper context.

For example, most treatments of star formation use relatively low values when implementing fixed local star formation efficiencies: \esf $ = 0.01 - 0.1$ \citep{Stinson2006,wang2015nihao}. As in our \texttt{rho0.1}, the density threshold is therefore not necessarily an accurate tracer of the actual density achieved by the gas. The actual distribution of gas density will depend more complexly on modeling prescriptions (i.e. realistic versus effective cooling treatments) when using \esf~$\ll 1$. For this reason, comparing simulations run with distinct modeling treatments but similar \rhoth~does not make sense when considering the dependence of core formation on \rhoth, as the resulting distribution of gas density and its sensitivity to feedback can vary substantially between models.



\section{Galaxy Structure}
\label{sec:structure}

Figure \ref{fig:10panelprojection} shows face-on and edge-on projections of the four alterations of the \texttt{SMUGGLE} model we consider, with the stellar half-mass radius (\rh) shown in green and \rcore~shown in magenta. The SH03 model shows a uniform disk with an unstructured ISM, along with large \rh~and small \rcore, while \texttt{fiducial} and \texttt{rho0.1} show a much more structured ISM, with clear fragmentation containing regions of both dense and rarefied gas. In addition, small pockets representing SN shock fronts can be seen in the face-on image. The disk remains well-behaved, with clear rotation and a roughly even distribution of gas throughout the disk. The ISM of \texttt{fiducial} is somewhat less evenly distributed than \texttt{rho0.1}, resulting in larger pockets of dense and rarefied gas, with an overall more centrally concentrated ISM (as seen in the edge-on projections), though it does maintain a disk morphology with clear cohesive rotation. Conversely, both \texttt{eSF100} and \texttt{vareff} have highly disturbed gaseous components with no clear rotation and strong radial outflows from more energetic SN feedback. Even the edge-on projections show little traditional disk structure, with the galaxies appearing irregular in structure. In addition, they are much more compact, with \rh~roughly half the size of those of SH03 or \texttt{rho0.1}. The core radii of the three \texttt{SMUGGLE} models are larger than that of the SH03 model (as shown previously).

\begin{figure}
    \centering
    \includegraphics[width=0.45\textwidth]{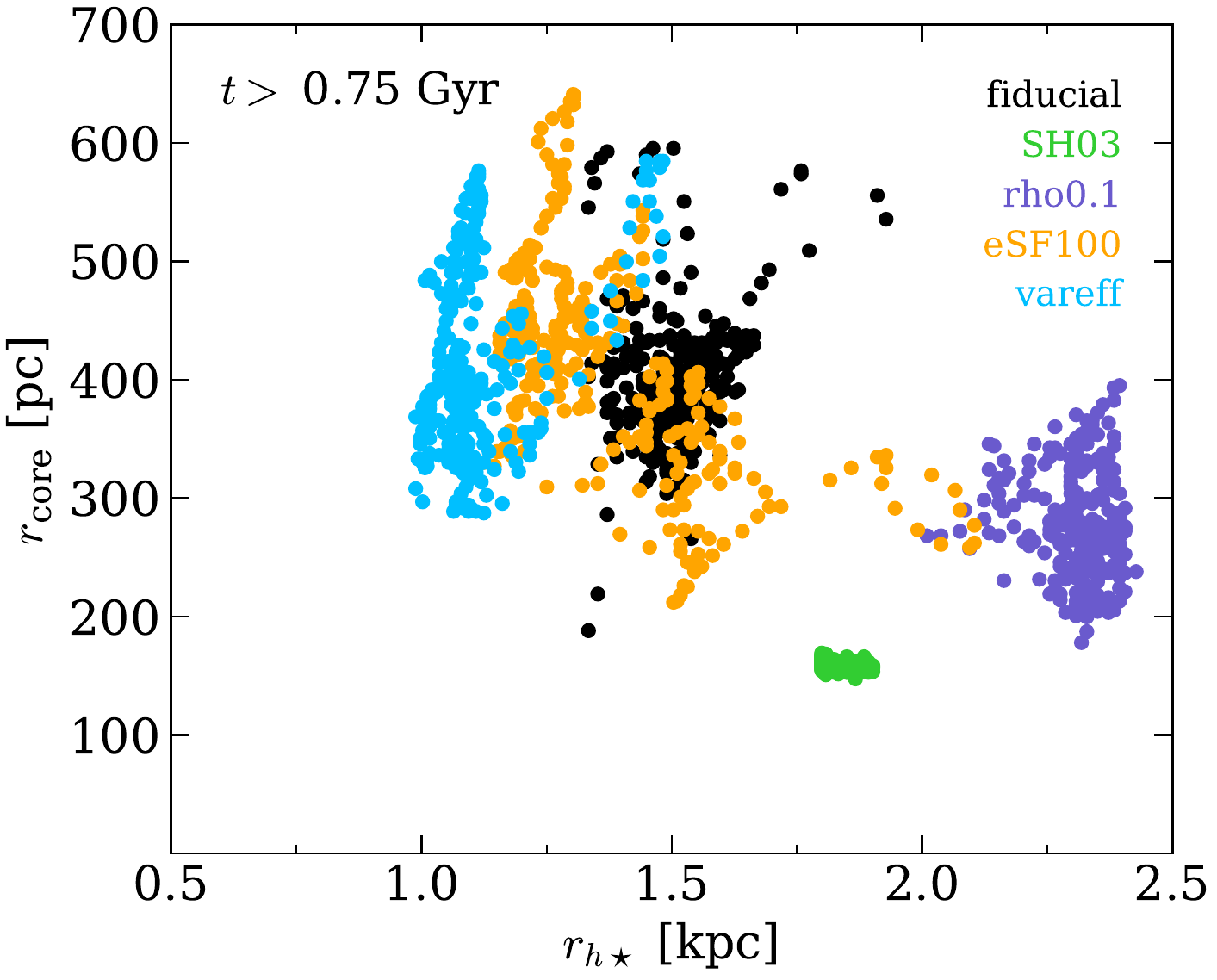}
    \caption{Core radius versus stellar half-mass radius for each run, with each point indicating a different snapshot after $t = 0.75$ Gyr. As in Figure \ref{fig:10panelprojection}, \rh~only includes star particles that were formed after the simulation began. Naturally, SH03 forms the tightest grouping, while the \texttt{SMUGGLE} runs are stratified according to galaxy size. It is a clear consequence of \texttt{vareff}'s prioritization of star formation in dense gas that it forms the most compact galaxy, while the global high efficiency of \texttt{eSF100} produces large fluctuations in galaxy size (and core radius). The low threshold of \texttt{rho0.1} allows for less dense gas in the outer regions to form stars, resulting in a more extended galaxy.}
    \label{fig:rcore_r50}
\end{figure}

\subsection{Morphology and cores}
Figure \ref{fig:rcore_r50} shows the core radius versus the half stellar mass radius for each run at $t > 0.75$ Gyr. We find a fair degree of stratification of the runs with \rh, indicating the effects of different modeling choices on galaxy structure. The variable efficiency run demonstrates the most compact galaxy size overall, mostly hovering aroung \rh~= 1 kpc. This concentrated morphology is a result of the maximized star formation efficiency in dense regions (which tend to be near the center of the galaxy) used in this model. The globally maximized star formation efficiency in \texttt{eSF100} produces a more concentrated galaxy than the fiducial \texttt{SMUGGLE} model, though it also has more variation. This run experienced a large burst of star formation at early times, expanding the initial galaxy, only contracting at later times. This expansion and contraction is seen in the orange dots that extend to the right of \rh~= 1.5 kpc, overlapping somewhat with our largest galaxy, \texttt{rho0.1}. 

Interestingly, the large core sizes and compact galaxies seen in \texttt{eSF100} and \texttt{vareff} are contrary to the observed trend in which large cores are expected in low surface-brightness galaxies \citep[][]{Santos-Santos2020}. This may indicate that cores can form in high surface brightness galaxies, but have not yet been detected (either due to incompleteness or the disruption of gas kinematics in systems that may mimic these runs), or it may indicate that high star formation efficiency is not an empirically consistent modeling choice. The latter may be more likely, since most ISM treatments that calibrate this parameter to observed data choose values in the range 0.01 -- 0.1 \citep{Stinson2006,wang2015nihao}, while models that implement such high efficiencies have other strict criteria on star formation \citep[][]{hopkins2014fire1}. Again, the effect of this parameter is indeed model-dependent. At least within \texttt{SMUGGLE}, an increased local SF efficiency parameter produces a trend counter to what is currently expected from observational data.

The large extent of \texttt{rho0.1} is a result of the reduced density threshold, which allows more rarefied gas in the outskirts of the galaxy to form stars, rather than concentrating star formation to the dense gas which collects near the center. The fiducial \texttt{SMUGGLE} model balances each of these effects, producing an intermediate-size galaxy, with \rh~$\approx 1.5$ kpc throughout its evolution. Each \texttt{SMUGGLE} model produces variation in both the core size and stellar half-mass radius. The SH03 model on the other hand maintains the same core radius and galaxy size throughout its evolution, forming a tight cluster of points. We note again that SH03 did not form a robust feedback-induced core. We include the data here only for contrast with our \texttt{SMUGGLE} runs which did form robust cores.

The variation in both core size and half stellar mass radius is worth noting. Observed galaxies can effectively only be measured at one point in their evolution. While a large sample of observed galaxies helps to sample the variation, it is still impossible to take into account the variation in these properties over a given galaxy's lifetime. It is certainly possible that extreme values of inner DM density from highly overdense cusps to underdense cores represent local maxima or minima in their fluctuations. We emphasize that a given observation is not necessarily representative of the property's expectation value. Numerically constraining the predicted fluctuation in such properties like DM core sizes may be a worthwhile addition to the discussion on diversity of rotation curves.

\begin{figure}
    \centering
    \includegraphics[width=0.45\textwidth]{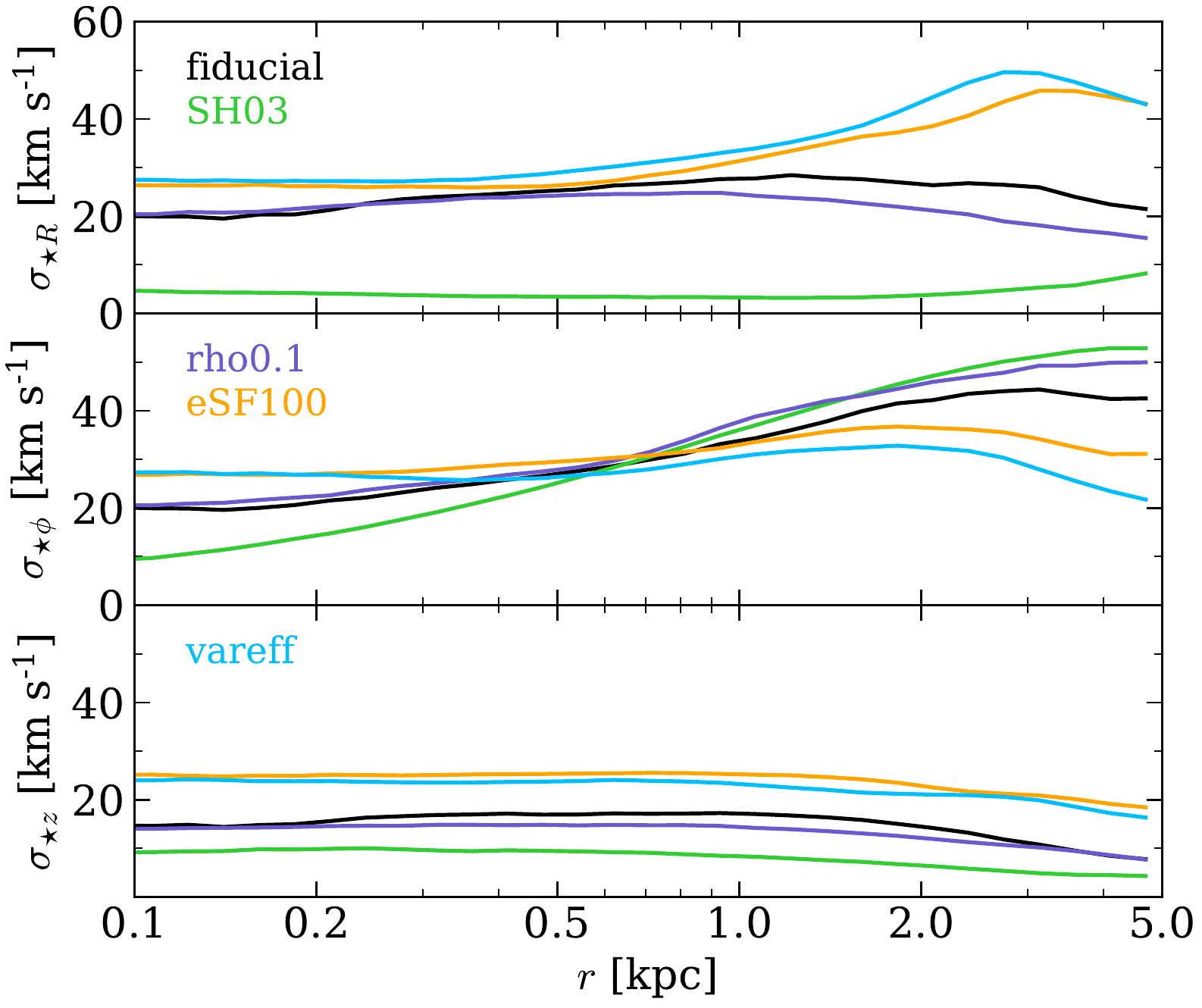}
    \caption{Time-averaged ($t > 0.75$ Gyr) stellar velocity dispersion $\sigma_\star$ in cylindrical coordinates. Standard errors are shown but appear smaller than the width of the lines. We find that SH03 preserves disk coherence better than the \texttt{SMUGGLE} models, which all produce stronger feedback that disrupts the rotational structure of the galaxy. The high efficiency variations distribute stellar motion more evenly between all three cylindrical components, indicating a dispersion-dominated galactic structure.}
    \label{fig:sigma}
\end{figure}

To quantify differences in the kinematic structure between our runs, Figure \ref{fig:sigma} shows time-averaged (for $t > 0.75$ Gyr) stellar velocity dispersion profiles in cylindrical coordinates, with $\sigma_R$ (radial direction) on the top panel, $\sigma_\phi$ (direction of disk rotation) in the center, and $\sigma_z$ (direction perpendicular to the disk plane) on the bottom panel. We see that all four \texttt{SMUGGLE} runs have higher $\sigma_R$ than SH03. The grouping of models echos that of their density distributions in Figure \ref{fig:rhohist1kpc_varISM}: \texttt{fiducial} and \texttt{rho0.1} have similar $\sigma_R$ profiles, and smaller than both \texttt{eSF100} and \texttt{vareff},  which are also similar to each other. This is a natural result of their higher star formation efficiencies, which results in stronger feedback, disturbing the ISM and causing increased radial motion into the gas due to increase SN activity. The center panel shows $\sigma_\phi$, representing the rotation of the disk. Disks with coherent rotation exhibit a typical ``S''-shaped curve, such as that of SH03, indicating a smooth increase in rotational velocity towards the outskirts of the galaxy. We see that \texttt{fiducial} and \texttt{rho0.1} exhibit this characteristic shape, but to a lesser degree as a result of their increased feedback. Naturally, the high efficiency models with their disrupted morphology show a near-constant $\sigma_\phi$ profile, indicating little to no rotational support. We observe a similar stratification of behavior in the bottom panel, where SH03 shows little gas motion in the $z$-direction, while \texttt{fiducial} and \texttt{rho0.1} show an increased amount, and \texttt{eSF100} and \texttt{vareff} show a stronger increase in gas disruption in this direction as a result of the strong feedback that injects a large amount of momentum in the local radial direction, resulting in increased gas velocity dispersions in all directions. Due to the broad similarity in core formation between the four \texttt{SMUGGLE} runs, this implies that the choice of star formation efficiency has little impact on the dark matter content while drastically affecting the gas content and morphology of dwarf galaxies.

\subsection{Diversity of rotation curves}
\label{sec:diversity}

\begin{figure}
    \centering
    \includegraphics[width=0.45\textwidth]{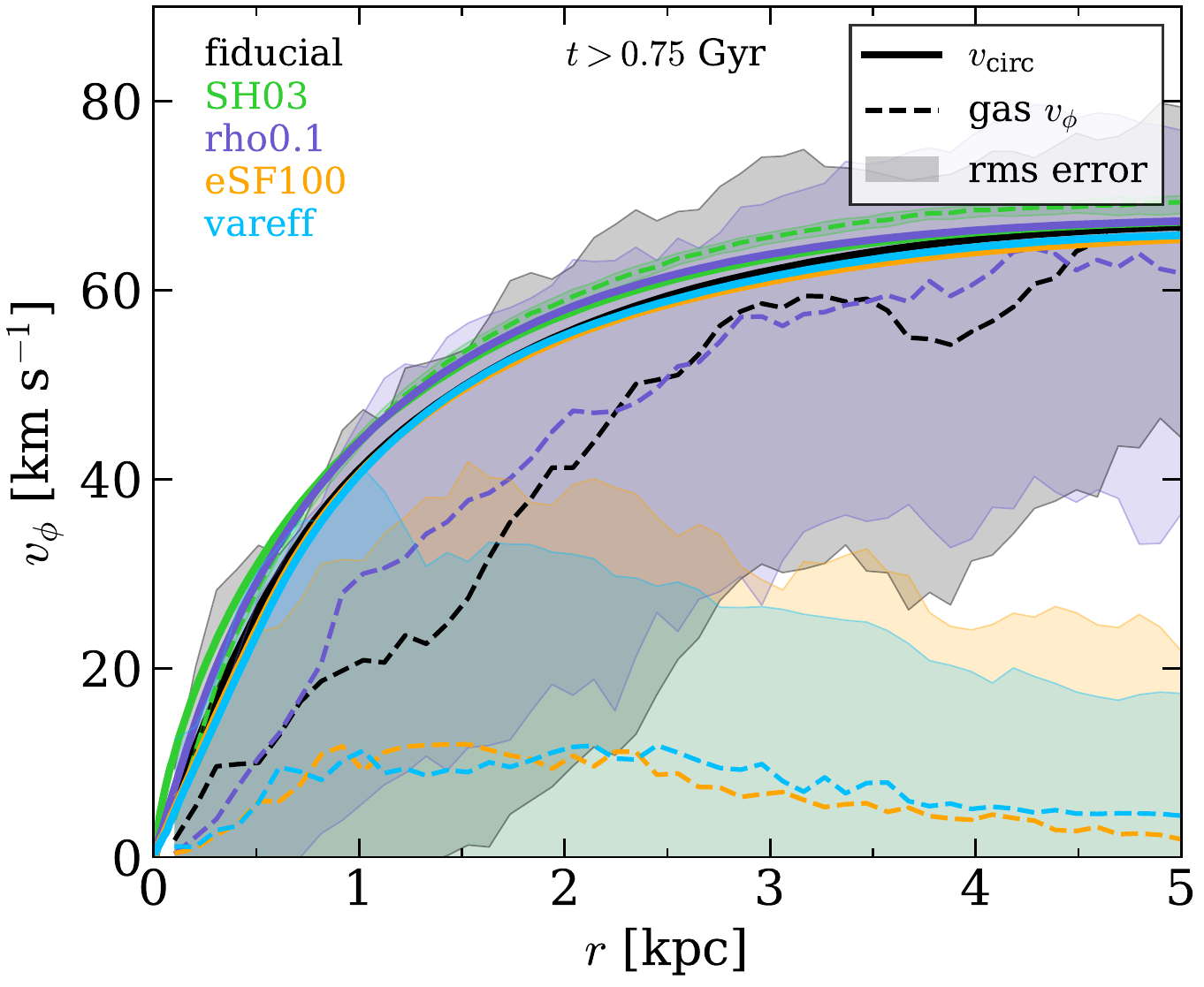}
    \caption{Median rotational velocity ($v_\phi$) profile of gas (dashed line) and total circular velocity \vcirc~= $\sqrt{\text{GM}(<r)/r}$ (solid line). Shaded regions represent the $1\sigma$ deviation from median (inner 68 per cent confidence interval) within each $r$ bin across time.
    We see that the (relatively) well-behaved ISM of \texttt{fiducial} and \texttt{rho0.1} trace the gravitational potential of the galaxy much better than the disturbed ISMs of \texttt{eSF100} and \texttt{vareff}. The large shaded errors indicate substantial variations in rotational velocity profiles over the course of the simulation.}
    \label{fig:vphi}
\end{figure}

Figure \ref{fig:vphi} shows the rotational velocity $v_\phi$ of the gas as well as \vcirc~= $\sqrt{\text{GM}(<r)/r}$ for each run, averaged over the final 0.5 Gyr of each run. We find that the ISM of SH03 traces the potential of the galaxy remarkably well. In contrast, the high efficiency \texttt{SMUGGLE} runs \texttt{eSF100} and \texttt{vareff} are so kinematically disrupted that there is little to no measurable rotation. \citet{elbadry2018b} found similarly dispersion-supported gas in dwarf galaxies within the FIRE simulations \citep{hopkins2018fire2}, and that rotational support increases with increasing mass. Further, they find that the majority of FIRE galaxies across 6.3 < $\log_{10}(M_\star / \text{M}_\odot)$ < 11.1 have little rotational support, and while the higher mass galaxies have morphological gas discs, only a fraction of the dwarf galaxies (\mstr~$\lesssim 10^9$ \msun) host this feature. 

It is notable that even within the `well-behaved' variations on \texttt{SMUGGLE}, we find that the rotational velocity of the gas does not accurately trace the \vcirc~implied by the gravitational potential. A naive reading of the gas $v_\phi$ distribution in Figure \ref{fig:vphi} could suggest a core radius of $\gtrsim$2 kpc for the fiducial \texttt{SMUGGLE} model and \texttt{rho0.1}, while our method of core size measurement relying only on DM density profiles (see Section \ref{sec:smuggle_and_SH03}) results in values of a few hundred parsecs. Interestingly, this 2 kpc figure is consistent with the fiducial radius used to compare well-resolved rotation curves between simulations and observations \citep[as in][]{Santos-Santos2018,Oman2019}. This result supports the notion that non-circular motion of gas in the inner regions of galaxies limits the ability of observational analyses to accurately recreate the DM profile, potentially contributing to the diversity of rotation curves in observed galaxies \citep{Oman2015,Oman2019,Santos-Santos2020}.

The variability in the  $v_\phi$ distribution indicates another problem of time sampling bias. The measured gas rotational velocity is subject to frequent and substantial variation as a result of energy injection via feedback, as depicted by the shaded regions on Figure \ref{fig:vphi} representing the RMS error due to time-averaging. Measurements of the $v_\phi$ distribution taken at the extrema of the error range could either produce rapidly rising rotation curves implying a mass distribution consistent with \lcdm, or a slowly rising rotation curve implying an inner mass deficit and substantial core. 

Figure \ref{fig:vphi_all} plots the individual $v_\phi$ profiles for each snapshot of the fiducial \texttt{SMUGGLE} run over the final 0.5 Gyr. Here we see that, while the majority of rotation curves are below the actual DM \vcirc, there are a handful of profiles that demonstrate rotation speeds faster than the DM in the inner regions, i.e. profiles that would be interpreted as cuspy. Based on the number of snapshots with rotation curves that rise faster than an NFW, we place an upper limit on the presence of highly cuspy rotation curves at approximately 10 per cent. While this is an unlikely result, it indicates that cuspy profiles as a result of gas kinematics are indeed possible.

The discrepancy between the rotational velocity $v_\phi$ profile and the circular velocity profile $v_\text{circ}$ indicates that the rotation of gas is rarely an accurate tracer of the DM potential in dwarf galaxies due to its sensitivity to energy injection via feedback. Our simulations predict that substantial diversity of rotation curves should be expected within the same dwarf galaxy across time. The variability of gas content and velocity in the inner regions of the galaxy on timescales $\lesssim 100$ Myr poses a challenge to the assumption of virial equilibrium (i.e. `steady-state') that underlies the inference of DM distributions from gas velocity profiles. As suggested by \citet{read2016rc}, expanding bubbles of HI can be used to identify post-starburst galaxies which are likely out of equilibrium. Collisionless stars may be a better tracer of the inner gravitational potential than gas. 

Overall, \texttt{SMUGGLE} produces rotational profiles that systematically underestimate the DM content of the inner regions, consistent with previous attempts to reconcile the observed diversity of rotation curves with baryonic physics \citep[][]{Oman2015,Oman2019,read2016rc,tollet2016,Santos-Santos2018,Santos-Santos2020}. This indicates either that our understanding of small-scale ISM physics within galaxies is incomplete, or that another mechanism is responsible for creating rapidly rising rotation curves. It is possible that higher mass systems with stronger potentials are less susceptible to this effect, but we emphasize that this must be demonstrated explicitly rather than taken as an assumption. 

The above considerations are only a result of ISM kinematics within an idealized, non-cosmological simulation and do not take into account additional bias introduced by observational measurement techniques, such as tilted-ring modeling and Jeans modeling, nor do they take into account evolutionary histories consistent with real galaxies or effects of cosmological environments such as mergers and infall of cold gas from filaments. Rather, these idealized tests isolate the effects of ISM modeling from other complex phenomena, allowing us to directly test the effects of baryonic feedback on the dark matter distribution of dwarf galaxies.

\section{Summary \& Conclusions}
\label{sec:summary}

We study the behavior of the \texttt{SMUGGLE} \citep{Marinacci2019} feedback and ISM model for the \textsc{Arepo} \citep{arepo} moving mesh simulation code. In particular, we investigate the formation of dark matter `cores' in idealized (non-cosmological) dwarf galaxies with \mstr~$\approx$ 8\e{7} \msun~and \mtwo~$\approx$ 2\e{10} \msun~by comparing runs with identical initial conditions under both \texttt{SMUGGLE} and the SH03 model \citep{SpringelHernquist2003}, a precursor to the model used in Illustris and Auriga \citep[][]{v14illustris,Grand2017} simulations, among others. We develop a self-consistent method of measuring the core radius to track its evolution over time. We define the core radius to be the location where the actual DM density falls below the predicted DM density of an NFW profile fit to the outer regions of the halo ($r > 3$ kpc) by a factor or 2 (Figure \ref{fig:rho_panel}). We then measure the slope of a power law fit to the resolved region within the measured core radius. Tracing these metrics over time, we find that SH03 does not produce a constant-density DM core, while the fiducial \texttt{SMUGGLE} model creates a flattened core of radius $\sim$ 350 pc within the first 0.75 Gyr. We show that the origin of these cores is linked to the successful self-regulation of the star formation history in \texttt{SMUGGLE} which establishes a bursty star formation mode. These bursty cycles then create significant variations in the enclosed gas mass within 1 kpc, resulting in non-adiabatic expansion of the inner DM distribution. Contrary to the self-regulation seen in \texttt{SMUGGLE}, SH03 produces a steadily declining SFH, with a constant mass of gas reached after most of the originally eligible gas for star formation has been transformed into stars. This equilibrium state then preserves the steep inner density profiles that have been reported previously in the literature (Figure \ref{fig:rcore_sfr}). 

In addition, we run three simulations of identical initial set up including alterations to key feedback parameters: (i) \texttt{rho0.1} changes the star formation density threshold from the fiducial value of \rhoth~$=100$ cm$^{-3}$ to a reduced value of \rhoth~$=0.1$ cm$^{-3}$; (ii) \texttt{eSF100} changes the local star formation efficiency (the mass fraction of eligible star forming gas that is converted into stars) from the fiducial value of \esf~$=0.01$ to an increased value of \esf~$=1$; and (iii) \texttt{vareff}, which implements a parameterization of the star formation efficiency based on the virial parameter (a measure of local self-gravitation; see Section \ref{sec:methods_variations}). We find that the formation of a core is robust to these changes in \texttt{SMUGGLE} (though \texttt{rho0.1} does form a $\sim 25$ per cent smaller core, and the high efficiency models exhibit stronger growth over time). 

It is significant that \texttt{rho0.1} forms a feedback-induced core while SH03 does not. Since both implementations use the same star formation density threshold \rhoth~$=0.1$ cm$^{-3}$, this is an indication that the density threshold alone is not a good predictor of core formation for detailed ISM models such as \texttt{SMUGGLE}. It is important to note that while SH03 does not generate a core through feedback, it does experience a halo expansion due to relaxation effects and the influence of the baryonic component \citep[][]{Burger2021}. Its expansion was smaller than in all \texttt{SMUGGLE} runs and was shown to be consistent with an adiabatic run, indicating that feedback was not a relevant factor. In contrast, \texttt{rho0.1} demonstrates large fluctuations of baryonic matter in the inner regions of the halo, linking feedback to core formation.

We find that the ability to resolve dense gas ($\rho_\text{gas} \gtrsim 10^2$ cm$^{-3}$; see Figure \ref{fig:rhohist1kpc_varISM}) is more predictive of core formation, in agreement with findings from \citep{benitezllambay2019}. For example, \texttt{rho0.1} resolves gas up to $\rho\sim 10^4$ cm$^{-3}$ while SH03 only resolves gas up to $\rho\sim 1$ cm$^{-3}$. This indicates that the SF density threshold is not a good proxy for actual gas density when using low local star formation efficiencies \esf~$\ll 1$. This then allows the dense gas to linger around and affect locally the gravitational potential even if the density threshold for star formation is nominally low. Note that this is different from predictions in other ISM implementations, such as NIHAO \citep{Dutton2019,Dutton2020}. 

Our high efficiency runs \texttt{eSF100} and \texttt{vareff} have more bursty star formation than \texttt{fiducial} or \texttt{rho0.1}, yet they do not form substantially larger cores (Figure \ref{fig:rcore_varISM}). This indicates that core size and burstiness are not tightly correlated, but that sufficiently bursty star formation, like sufficiently high gas density, are necessary conditions for core formation, as predicted previously \citep{pontzengovernato2012,benitezllambay2019,Dutton2019}. All \texttt{SMUGGLE} variations also demonstrate mild time-dependence over the course of our runs, indicating that core expansion should continue over cosmological timescales. We hypothesize that the source of this continued expansion is the continued bursty star formation in these runs. The core evolution in the fiducial \texttt{SMUGGLE} run is inconclusive in its time-dependence due to the short runtime of these simulations. Density profiles of the \texttt{SMUGGLE} variations can be found in Figure \ref{fig:fourrho_varISM}.

While there is broad agreement in core formation between the \texttt{SMUGGLE} variations, there are still differences between the models: \texttt{rho0.1} forms the smallest core of the \texttt{SMUGGLE} models, with \rcore $\sim$300 pc by final time, while \texttt{vareff} and \texttt{eSF100} reach final core radii of $\sim$500 pc. Despite this difference, we maintain that \texttt{rho0.1} does indeed form a comparable core due to its highly flattened inner slope consistent with the fiducial \texttt{SMUGGLE} model. Interestingly, the fluctuations in gas mass within 1 kpc for all \texttt{SMUGGLE} runs are comparable (though with \texttt{rho0.1} having less frequent outflows). This is likely the source of the similarity in core sizes and shapes between the runs.

This similarity between variations of \texttt{SMUGGLE} indicates that the physical consequences of changing parameters such as the SF density threshold \rhoth~are highly model dependent. As mentioned, \rhoth~is not an accurate tracer of physical gas densities achieved by simulations when using empirically calibrated models that limit the local SF efficiency \esf~to values $\ll 1$. Local gas densities will be highly dependent on implementations of subgrid physics. In particular, molecular and fine-structure cooling allows gas to naturally reach temperatures far lower than $10^4$ K and achieve densities comparable to or higher than the average density of DM in the inner regions. The implemented modes of feedback-driven energy injection into the ISM allow this dense gas to be disrupted and flow to outer regions of the halo, repeatedly perturbing the DM potential as suggested by \citet{pontzengovernato2012}. That is to say, changes in model parameters must result in the required physical changes within the simulation to accurately capture the details of baryon-induced core formation. Simulations that do not produce sufficiently dense gas (due either to modeling choices or resolution) are simply unable to follow the physics expected to lead to core formation.

We also investigate the implications various modeling choices have on morphology. The fiducial \texttt{SMUGGLE} model and \texttt{rho0.1} both form rotationally supported disks with structured ISMs, while SH03 naturally produces a stable galaxy with featureless ISM (see Figures \ref{fig:10panelprojection} and \ref{fig:sigma}). On the other hand, the high efficiency models produce dispersion-dominated spheroid galaxies with lower gas fractions. This is a natural result of the increased burstiness and feedback of these models, and is in agreement with the FIRE simulations \citep{hopkins2018fire2}, which implement \esf $=1$ and also observe dwarf galaxies with spheroid morphology and dispersion supported ISM \citep{elbadry2018b}. Interestingly, we find that the most compact galaxies (\texttt{eSF100} and \texttt{vareff}) form the largest cores, while the most diffuse galaxies (\texttt{rho0.1}) form the smallest cores (Figure \ref{fig:rcore_r50}), in agreement with \citet{Burger2021}. 

Our examination of the rotational velocity ($v_\phi$) profiles of the gas content (Figure \ref{fig:vphi}) indicates that the ISM does not trace the potential of the DM in the inner regions ($r < 2$ kpc). This is true for all \texttt{SMUGGLE} variations, though the fiducial model and \texttt{rho0.1} are better able to trace the DM \vcirc~in the outskirts, while \texttt{eSF100} and \texttt{vareff} demonstrate almost no rotational velocity component of the gas at any radius. Further, we find significant variations in the $v_\phi$ profiles across time, suggesting that a diverse morphology of rotation curves can be observed at different times within the same galaxy. We find that individual $v_\phi$ profiles can vary between exceeding the expected DM circular velocity and drastically underestimating it (Figure \ref{fig:vphi_all}). However, we find that the ISM in \texttt{SMUGGLE} systematically falls below the \vcirc~of the halo within the inner regions, consistent with previous work \citep[][]{Santos-Santos2020,Oman2019}, further indicating that the baryon-induced core formation scheme struggles to reproduce the steep end of the diversity of rotation curves problem.

Our analysis indicates that feedback-induced core formation is fundamentally a small-scale problem. Its effects may be observed on the scale of a few kpc, but the physics which generates these observables occur on the scales of star formation and feedback, i.e. $10 - 100$ pc, as well as sub-pc processes that are yet unresolved and only implemented through sub-grid modeling. Lack of cores in models which are not able (and do not attempt) to produce this microphysics is not evidence against the validity of baryon-induced core formation, but evidence against the suitability of such models to study this process. 

Finally, our results suggest that even if perfect observations of gas rotation curves are obtained, these do not necessarily trace the DM potential in non-equilibrium systems such as dispersion-dominated dwarf galaxies. Caution is needed when attempting to infer DM distributions from gas rotation. It is important to investigate the assumption of equilibrium for dwarf galaxies, whose small sizes make them susceptible to large fluctuations in gas content and velocity. 


\section*{Acknowledgements}

EDJ and LVS acknowledge support from the NASA ATP 80NSSC20K0566, NSF AST 1817233 and NSF CAREER 1945310 grants. 
FM acknowledges support through the program "Rita Levi Montalcini" of the Italian MIUR.
MV acknowledges support through NASA ATP grants 16-ATP16-0167, 19-ATP19-0019, 19-ATP19-0020, 19-ATP19-0167, and NSF grants AST-1814053, AST-1814259, AST-1909831, AST-2007355 and AST-2107724.
PT and JQ acknowledge support from NSF grants AST-200849. PT acknowledges support from NASA ATP grant 80NSSC20K0502.
AS and HL acknowledge support for Program numbers HST-HF2-51421.001-A and HST-HF2-51438.001-A provided by NASA through a grant from the Space Telescope Science Institute,
which is operated by the Association of Universities for Research in Astronomy, Incorporated, under NASA contract NAS5-26555. 
JB acknowledges support by a Grant of Excellence from the Icelandic Center for Research (Rann\'is; grant number 173929)
JZ acknowledges support by a Grant of Excellence from the Icelandic Research fund (grant number 206930). 


\section*{Data Availability}
The data that support the findings of this study are available from the corresponding author, upon reasonable request.




\bibliographystyle{mnras}
\bibliography{cores} 

\begin{thebibliography}{}
\makeatletter
\relax
\def\mn@urlcharsother{\let\do\@makeother \do\$\do\&\do\#\do\^\do\_\do\%\do\~}
\def\mn@doi{\begingroup\mn@urlcharsother \@ifnextchar [ {\mn@doi@}
  {\mn@doi@[]}}
\def\mn@doi@[#1]#2{\def\@tempa{#1}\ifx\@tempa\@empty \href
  {http://dx.doi.org/#2} {doi:#2}\else \href {http://dx.doi.org/#2} {#1}\fi
  \endgroup}
\def\mn@eprint#1#2{\mn@eprint@#1:#2::\@nil}
\def\mn@eprint@arXiv#1{\href {http://arxiv.org/abs/#1} {{\tt arXiv:#1}}}
\def\mn@eprint@dblp#1{\href {http://dblp.uni-trier.de/rec/bibtex/#1.xml}
  {dblp:#1}}
\def\mn@eprint@#1:#2:#3:#4\@nil{\def\@tempa {#1}\def\@tempb {#2}\def\@tempc
  {#3}\ifx \@tempc \@empty \let \@tempc \@tempb \let \@tempb \@tempa \fi \ifx
  \@tempb \@empty \def\@tempb {arXiv}\fi \@ifundefined
  {mn@eprint@\@tempb}{\@tempb:\@tempc}{\expandafter \expandafter \csname
  mn@eprint@\@tempb\endcsname \expandafter{\@tempc}}}

\bibitem[\protect\citeauthoryear{{Adams} et~al.,}{{Adams}
  et~al.}{2014}]{Adams2014}
{Adams} J.~J.,  et~al., 2014, \mn@doi [\apj] {10.1088/0004-637X/789/1/63},
  \href {https://ui.adsabs.harvard.edu/abs/2014ApJ...789...63A} {789, 63}

\bibitem[\protect\citeauthoryear{{Ben{\'\i}tez-Llambay}, {Frenk}, {Ludlow}  \&
  {Navarro}}{{Ben{\'\i}tez-Llambay} et~al.}{2019}]{benitezllambay2019}
{Ben{\'\i}tez-Llambay} A.,  {Frenk} C.~S.,  {Ludlow} A.~D.,   {Navarro} J.~F.,
  2019, \mn@doi [\mnras] {10.1093/mnras/stz1890}, \href
  {https://ui.adsabs.harvard.edu/abs/2019MNRAS.488.2387B} {488, 2387}

\bibitem[\protect\citeauthoryear{{Bode}, {Ostriker}  \& {Turok}}{{Bode}
  et~al.}{2001}]{Bode2001}
{Bode} P.,  {Ostriker} J.~P.,   {Turok} N.,  2001, \mn@doi [\apj]
  {10.1086/321541}, \href
  {https://ui.adsabs.harvard.edu/abs/2001ApJ...556...93B} {556, 93}

\bibitem[\protect\citeauthoryear{{Bose} et~al.,}{{Bose}
  et~al.}{2019}]{bose2019}
{Bose} S.,  et~al., 2019, \mn@doi [\mnras] {10.1093/mnras/stz1168}, \href
  {https://ui.adsabs.harvard.edu/abs/2019MNRAS.486.4790B} {486, 4790}

\bibitem[\protect\citeauthoryear{{Bozek} et~al.,}{{Bozek}
  et~al.}{2019}]{Bozek2019}
{Bozek} B.,  et~al., 2019, \mn@doi [\mnras] {10.1093/mnras/sty3300}, \href
  {https://ui.adsabs.harvard.edu/abs/2019MNRAS.483.4086B} {483, 4086}

\bibitem[\protect\citeauthoryear{{Brooks} \& {Zolotov}}{{Brooks} \&
  {Zolotov}}{2014}]{Brooks2014}
{Brooks} A.~M.,  {Zolotov} A.,  2014, \mn@doi [\apj]
  {10.1088/0004-637X/786/2/87}, \href
  {https://ui.adsabs.harvard.edu/abs/2014ApJ...786...87B} {786, 87}

\bibitem[\protect\citeauthoryear{{Burger} \& {Zavala}}{{Burger} \&
  {Zavala}}{2019}]{Burger2019}
{Burger} J.~D.,  {Zavala} J.,  2019, \mn@doi [\mnras] {10.1093/mnras/stz496},
  \href {https://ui.adsabs.harvard.edu/abs/2019MNRAS.485.1008B} {485, 1008}

\bibitem[\protect\citeauthoryear{{Burger} \& {Zavala}}{{Burger} \&
  {Zavala}}{2021}]{Burger2021}
{Burger} J.~D.,  {Zavala} J.,  2021, arXiv e-prints, \href
  {https://ui.adsabs.harvard.edu/abs/2021arXiv210301231B} {p. arXiv:2103.01231}

\bibitem[\protect\citeauthoryear{{Burkert}}{{Burkert}}{1995}]{Burkert1995}
{Burkert} A.,  1995, \mn@doi [\apjl] {10.1086/309560}, \href
  {https://ui.adsabs.harvard.edu/abs/1995ApJ...447L..25B} {447, L25}

\bibitem[\protect\citeauthoryear{{Burkert}}{{Burkert}}{2020}]{Burkert2020}
{Burkert} A.,  2020, \mn@doi [\apj] {10.3847/1538-4357/abb242}, \href
  {https://ui.adsabs.harvard.edu/abs/2020ApJ...904..161B} {904, 161}

\bibitem[\protect\citeauthoryear{{Chabrier}}{{Chabrier}}{2001}]{chabrier2001}
{Chabrier} G.,  2001, \mn@doi [\apj] {10.1086/321401}, \href
  {https://ui.adsabs.harvard.edu/abs/2001ApJ...554.1274C} {554, 1274}

\bibitem[\protect\citeauthoryear{{Chan}, {Kere{\v{s}}}, {O{\~n}orbe},
  {Hopkins}, {Muratov}, {Faucher-Gigu{\`e}re}  \& {Quataert}}{{Chan}
  et~al.}{2015}]{chan2015}
{Chan} T.~K.,  {Kere{\v{s}}} D.,  {O{\~n}orbe} J.,  {Hopkins} P.~F.,  {Muratov}
  A.~L.,  {Faucher-Gigu{\`e}re} C.~A.,   {Quataert} E.,  2015, \mn@doi [\mnras]
  {10.1093/mnras/stv2165}, \href
  {https://ui.adsabs.harvard.edu/abs/2015MNRAS.454.2981C} {454, 2981}

\bibitem[\protect\citeauthoryear{{Chua}, {Pillepich}, {Vogelsberger}  \&
  {Hernquist}}{{Chua} et~al.}{2019}]{Chua2019}
{Chua} K. T.~E.,  {Pillepich} A.,  {Vogelsberger} M.,   {Hernquist} L.,  2019,
  \mn@doi [\mnras] {10.1093/mnras/sty3531}, \href
  {https://ui.adsabs.harvard.edu/abs/2019MNRAS.484..476C} {484, 476}

\bibitem[\protect\citeauthoryear{{Cioffi}, {McKee}  \& {Bertschinger}}{{Cioffi}
  et~al.}{1988}]{Cioffi1988}
{Cioffi} D.~F.,  {McKee} C.~F.,   {Bertschinger} E.,  1988, \mn@doi [\apj]
  {10.1086/166834}, \href
  {https://ui.adsabs.harvard.edu/abs/1988ApJ...334..252C} {334, 252}

\bibitem[\protect\citeauthoryear{{Crain} et~al.,}{{Crain}
  et~al.}{2015}]{crain2015}
{Crain} R.~A.,  et~al., 2015, \mn@doi [\mnras] {10.1093/mnras/stv725}, \href
  {https://ui.adsabs.harvard.edu/abs/2015MNRAS.450.1937C} {450, 1937}

\bibitem[\protect\citeauthoryear{{Creasey}, {Sameie}, {Sales}, {Yu},
  {Vogelsberger}  \& {Zavala}}{{Creasey} et~al.}{2017}]{Creasey2017}
{Creasey} P.,  {Sameie} O.,  {Sales} L.~V.,  {Yu} H.-B.,  {Vogelsberger} M.,
  {Zavala} J.,  2017, \mn@doi [\mnras] {10.1093/mnras/stx522}, \href
  {https://ui.adsabs.harvard.edu/abs/2017MNRAS.468.2283C} {468, 2283}

\bibitem[\protect\citeauthoryear{{Dav{\'e}}, {Oppenheimer}  \&
  {Finlator}}{{Dav{\'e}} et~al.}{2011}]{Dave2011}
{Dav{\'e}} R.,  {Oppenheimer} B.~D.,   {Finlator} K.,  2011, \mn@doi [\mnras]
  {10.1111/j.1365-2966.2011.18680.x}, \href
  {https://ui.adsabs.harvard.edu/abs/2011MNRAS.415...11D} {415, 11}

\bibitem[\protect\citeauthoryear{{Dav{\'e}}, {Angl{\'e}s-Alc{\'a}zar},
  {Narayanan}, {Li}, {Rafieferantsoa}  \& {Appleby}}{{Dav{\'e}}
  et~al.}{2019}]{Dave2019}
{Dav{\'e}} R.,  {Angl{\'e}s-Alc{\'a}zar} D.,  {Narayanan} D.,  {Li} Q.,
  {Rafieferantsoa} M.~H.,   {Appleby} S.,  2019, \mn@doi [\mnras]
  {10.1093/mnras/stz937}, \href
  {https://ui.adsabs.harvard.edu/abs/2019MNRAS.486.2827D} {486, 2827}

\bibitem[\protect\citeauthoryear{{Di Cintio}, {Brook}, {Macci{\`o}}, {Stinson},
  {Knebe}, {Dutton}  \& {Wadsley}}{{Di Cintio} et~al.}{2014}]{dicintio2014SMHM}
{Di Cintio} A.,  {Brook} C.~B.,  {Macci{\`o}} A.~V.,  {Stinson} G.~S.,  {Knebe}
  A.,  {Dutton} A.~A.,   {Wadsley} J.,  2014, \mn@doi [\mnras]
  {10.1093/mnras/stt1891}, \href
  {https://ui.adsabs.harvard.edu/abs/2014MNRAS.437..415D} {437, 415}

\bibitem[\protect\citeauthoryear{{Dodelson} \& {Widrow}}{{Dodelson} \&
  {Widrow}}{1994}]{Dodelson1994}
{Dodelson} S.,  {Widrow} L.~M.,  1994, \mn@doi [\prl]
  {10.1103/PhysRevLett.72.17}, \href
  {https://ui.adsabs.harvard.edu/abs/1994PhRvL..72...17D} {72, 17}

\bibitem[\protect\citeauthoryear{{Dubois} et~al.,}{{Dubois}
  et~al.}{2014}]{Dubois2014}
{Dubois} Y.,  et~al., 2014, \mn@doi [\mnras] {10.1093/mnras/stu1227}, \href
  {https://ui.adsabs.harvard.edu/abs/2014MNRAS.444.1453D} {444, 1453}

\bibitem[\protect\citeauthoryear{{Dutton} et~al.,}{{Dutton}
  et~al.}{2016}]{Dutton2016}
{Dutton} A.~A.,  et~al., 2016, \mn@doi [\mnras] {10.1093/mnras/stw1537}, \href
  {https://ui.adsabs.harvard.edu/abs/2016MNRAS.461.2658D} {461, 2658}

\bibitem[\protect\citeauthoryear{{Dutton}, {Macci{\`o}}, {Buck}, {Dixon},
  {Blank}  \& {Obreja}}{{Dutton} et~al.}{2019}]{Dutton2019}
{Dutton} A.~A.,  {Macci{\`o}} A.~V.,  {Buck} T.,  {Dixon} K.~L.,  {Blank} M.,
  {Obreja} A.,  2019, \mn@doi [\mnras] {10.1093/mnras/stz889}, \href
  {https://ui.adsabs.harvard.edu/abs/2019MNRAS.486..655D} {486, 655}

\bibitem[\protect\citeauthoryear{{Dutton}, {Buck}, {Macci{\`o}}, {Dixon},
  {Blank}  \& {Obreja}}{{Dutton} et~al.}{2020}]{Dutton2020}
{Dutton} A.~A.,  {Buck} T.,  {Macci{\`o}} A.~V.,  {Dixon} K.~L.,  {Blank} M.,
  {Obreja} A.,  2020, \mn@doi [\mnras] {10.1093/mnras/staa3028}, \href
  {https://ui.adsabs.harvard.edu/abs/2020MNRAS.499.2648D} {499, 2648}

\bibitem[\protect\citeauthoryear{{El-Badry} et~al.,}{{El-Badry}
  et~al.}{2018}]{elbadry2018b}
{El-Badry} K.,  et~al., 2018, \mn@doi [\mnras] {10.1093/mnras/stx2482}, \href
  {https://ui.adsabs.harvard.edu/abs/2018MNRAS.473.1930E} {473, 1930}

\bibitem[\protect\citeauthoryear{{Elbert}, {Bullock}, {Kaplinghat},
  {Garrison-Kimmel}, {Graus}  \& {Rocha}}{{Elbert} et~al.}{2018}]{Elbert2018}
{Elbert} O.~D.,  {Bullock} J.~S.,  {Kaplinghat} M.,  {Garrison-Kimmel} S.,
  {Graus} A.~S.,   {Rocha} M.,  2018, \mn@doi [\apj]
  {10.3847/1538-4357/aa9710}, \href
  {https://ui.adsabs.harvard.edu/abs/2018ApJ...853..109E} {853, 109}

\bibitem[\protect\citeauthoryear{{Faucher-Gigu{\`e}re}, {Lidz}, {Zaldarriaga}
  \& {Hernquist}}{{Faucher-Gigu{\`e}re} et~al.}{2009}]{faucher2009}
{Faucher-Gigu{\`e}re} C.-A.,  {Lidz} A.,  {Zaldarriaga} M.,   {Hernquist} L.,
  2009, \mn@doi [\apj] {10.1088/0004-637X/703/2/1416}, \href
  {https://ui.adsabs.harvard.edu/abs/2009ApJ...703.1416F} {703, 1416}

\bibitem[\protect\citeauthoryear{{Ferri{\`e}re}}{{Ferri{\`e}re}}{2001}]{Ferriere2001}
{Ferri{\`e}re} K.~M.,  2001, \mn@doi [Reviews of Modern Physics]
  {10.1103/RevModPhys.73.1031}, \href
  {https://ui.adsabs.harvard.edu/abs/2001RvMP...73.1031F} {73, 1031}

\bibitem[\protect\citeauthoryear{{Fitts} et~al.,}{{Fitts}
  et~al.}{2017}]{fitts2017}
{Fitts} A.,  et~al., 2017, \mn@doi [\mnras] {10.1093/mnras/stx1757}, \href
  {https://ui.adsabs.harvard.edu/abs/2017MNRAS.471.3547F} {471, 3547}

\bibitem[\protect\citeauthoryear{{Fitts} et~al.,}{{Fitts}
  et~al.}{2019}]{Fitts2019}
{Fitts} A.,  et~al., 2019, \mn@doi [\mnras] {10.1093/mnras/stz2613}, \href
  {https://ui.adsabs.harvard.edu/abs/2019MNRAS.490..962F} {490, 962}

\bibitem[\protect\citeauthoryear{{Flores} \& {Primack}}{{Flores} \&
  {Primack}}{1994}]{FloresPrimack1994}
{Flores} R.~A.,  {Primack} J.~R.,  1994, \mn@doi [\apjl] {10.1086/187350},
  \href {https://ui.adsabs.harvard.edu/abs/1994ApJ...427L...1F} {427, L1}

\bibitem[\protect\citeauthoryear{{Genel} et~al.,}{{Genel}
  et~al.}{2014}]{Genel2014}
{Genel} S.,  et~al., 2014, \mn@doi [\mnras] {10.1093/mnras/stu1654}, \href
  {https://ui.adsabs.harvard.edu/abs/2014MNRAS.445..175G} {445, 175}

\bibitem[\protect\citeauthoryear{{Genina} et~al.,}{{Genina}
  et~al.}{2018}]{Genina2018}
{Genina} A.,  et~al., 2018, \mn@doi [\mnras] {10.1093/mnras/stx2855}, \href
  {https://ui.adsabs.harvard.edu/abs/2018MNRAS.474.1398G} {474, 1398}

\bibitem[\protect\citeauthoryear{{Gilmore}, {Wilkinson}, {Wyse}, {Kleyna},
  {Koch}, {Evans}  \& {Grebel}}{{Gilmore} et~al.}{2007}]{Gilmore2007}
{Gilmore} G.,  {Wilkinson} M.~I.,  {Wyse} R. F.~G.,  {Kleyna} J.~T.,  {Koch}
  A.,  {Evans} N.~W.,   {Grebel} E.~K.,  2007, \mn@doi [\apj] {10.1086/518025},
  \href {https://ui.adsabs.harvard.edu/abs/2007ApJ...663..948G} {663, 948}

\bibitem[\protect\citeauthoryear{{Governato} et~al.,}{{Governato}
  et~al.}{2010}]{Governato2010}
{Governato} F.,  et~al., 2010, \mn@doi [\nat] {10.1038/nature08640}, \href
  {https://ui.adsabs.harvard.edu/abs/2010Natur.463..203G} {463, 203}

\bibitem[\protect\citeauthoryear{{Grand} et~al.,}{{Grand}
  et~al.}{2017}]{Grand2017}
{Grand} R. J.~J.,  et~al., 2017, \mn@doi [\mnras] {10.1093/mnras/stx071}, \href
  {https://ui.adsabs.harvard.edu/abs/2017MNRAS.467..179G} {467, 179}

\bibitem[\protect\citeauthoryear{{Greggio}}{{Greggio}}{2005}]{greggio2005}
{Greggio} L.,  2005, \mn@doi [\aap] {10.1051/0004-6361:20052926}, \href
  {https://ui.adsabs.harvard.edu/abs/2005A&A...441.1055G} {441, 1055}

\bibitem[\protect\citeauthoryear{{Hernquist}}{{Hernquist}}{1990}]{Hernquist1990}
{Hernquist} L.,  1990, \mn@doi [\apj] {10.1086/168845}, \href
  {https://ui.adsabs.harvard.edu/abs/1990ApJ...356..359H} {356, 359}

\bibitem[\protect\citeauthoryear{{Hopkins}, {Quataert}  \& {Murray}}{{Hopkins}
  et~al.}{2011}]{Hopkins2011}
{Hopkins} P.~F.,  {Quataert} E.,   {Murray} N.,  2011, \mn@doi [\mnras]
  {10.1111/j.1365-2966.2011.19306.x}, \href
  {https://ui.adsabs.harvard.edu/abs/2011MNRAS.417..950H} {417, 950}

\bibitem[\protect\citeauthoryear{{Hopkins}, {Kere{\v{s}}}, {O{\~n}orbe},
  {Faucher-Gigu{\`e}re}, {Quataert}, {Murray}  \& {Bullock}}{{Hopkins}
  et~al.}{2014}]{hopkins2014fire1}
{Hopkins} P.~F.,  {Kere{\v{s}}} D.,  {O{\~n}orbe} J.,  {Faucher-Gigu{\`e}re}
  C.-A.,  {Quataert} E.,  {Murray} N.,   {Bullock} J.~S.,  2014, \mn@doi
  [\mnras] {10.1093/mnras/stu1738}, \href
  {https://ui.adsabs.harvard.edu/abs/2014MNRAS.445..581H} {445, 581}

\bibitem[\protect\citeauthoryear{{Hopkins} et~al.,}{{Hopkins}
  et~al.}{2018}]{hopkins2018fire2}
{Hopkins} P.~F.,  et~al., 2018, \mn@doi [\mnras] {10.1093/mnras/sty1690}, \href
  {https://ui.adsabs.harvard.edu/abs/2018MNRAS.480..800H} {480, 800}

\bibitem[\protect\citeauthoryear{{Hu}, {Barkana}  \& {Gruzinov}}{{Hu}
  et~al.}{2000}]{Hu2000}
{Hu} W.,  {Barkana} R.,   {Gruzinov} A.,  2000, \mn@doi [\prl]
  {10.1103/PhysRevLett.85.1158}, \href
  {https://ui.adsabs.harvard.edu/abs/2000PhRvL..85.1158H} {85, 1158}

\bibitem[\protect\citeauthoryear{{Ikeuchi} \& {Ostriker}}{{Ikeuchi} \&
  {Ostriker}}{1986}]{Ikeuchi1986}
{Ikeuchi} S.,  {Ostriker} J.~P.,  1986, \mn@doi [\apj] {10.1086/163921}, \href
  {https://ui.adsabs.harvard.edu/abs/1986ApJ...301..522I} {301, 522}

\bibitem[\protect\citeauthoryear{{Iorio}, {Fraternali}, {Nipoti}, {Di Teodoro},
  {Read}  \& {Battaglia}}{{Iorio} et~al.}{2017}]{Iorio2017}
{Iorio} G.,  {Fraternali} F.,  {Nipoti} C.,  {Di Teodoro} E.,  {Read} J.~I.,
  {Battaglia} G.,  2017, \mn@doi [\mnras] {10.1093/mnras/stw3285}, \href
  {https://ui.adsabs.harvard.edu/abs/2017MNRAS.466.4159I} {466, 4159}

\bibitem[\protect\citeauthoryear{{Kannan}, {Marinacci}, {Vogelsberger},
  {Sales}, {Torrey}, {Springel}  \& {Hernquist}}{{Kannan}
  et~al.}{2020}]{Kannan2020}
{Kannan} R.,  {Marinacci} F.,  {Vogelsberger} M.,  {Sales} L.~V.,  {Torrey} P.,
   {Springel} V.,   {Hernquist} L.,  2020, \mn@doi [\mnras]
  {10.1093/mnras/staa3249}, \href
  {https://ui.adsabs.harvard.edu/abs/2020MNRAS.499.5732K} {499, 5732}

\bibitem[\protect\citeauthoryear{{Kaplinghat}, {Ren}  \& {Yu}}{{Kaplinghat}
  et~al.}{2020}]{Kaplinghat2020}
{Kaplinghat} M.,  {Ren} T.,   {Yu} H.-B.,  2020, \mn@doi [\jcap]
  {10.1088/1475-7516/2020/06/027}, \href
  {https://ui.adsabs.harvard.edu/abs/2020JCAP...06..027K} {2020, 027}

\bibitem[\protect\citeauthoryear{{Klypin}, {Kravtsov}, {Bullock}  \&
  {Primack}}{{Klypin} et~al.}{2001}]{Klypin2001}
{Klypin} A.,  {Kravtsov} A.~V.,  {Bullock} J.~S.,   {Primack} J.~R.,  2001,
  \mn@doi [\apj] {10.1086/321400}, \href
  {https://ui.adsabs.harvard.edu/abs/2001ApJ...554..903K} {554, 903}

\bibitem[\protect\citeauthoryear{{Kormendy}, {Fisher}, {Cornell}  \&
  {Bender}}{{Kormendy} et~al.}{2009}]{Kormendy2009}
{Kormendy} J.,  {Fisher} D.~B.,  {Cornell} M.~E.,   {Bender} R.,  2009, \mn@doi
  [\apjs] {10.1088/0067-0049/182/1/216}, \href
  {https://ui.adsabs.harvard.edu/abs/2009ApJS..182..216K} {182, 216}

\bibitem[\protect\citeauthoryear{{Kuzio de Naray}, {McGaugh}, {de Blok}  \&
  {Bosma}}{{Kuzio de Naray} et~al.}{2006}]{KuzioDeNaray2006}
{Kuzio de Naray} R.,  {McGaugh} S.~S.,  {de Blok} W.~J.~G.,   {Bosma} A.,
  2006, \mn@doi [\apjs] {10.1086/505345}, \href
  {https://ui.adsabs.harvard.edu/abs/2006ApJS..165..461K} {165, 461}

\bibitem[\protect\citeauthoryear{{Kuzio de Naray}, {McGaugh}  \& {de
  Blok}}{{Kuzio de Naray} et~al.}{2008}]{KuzioDeNaray2008}
{Kuzio de Naray} R.,  {McGaugh} S.~S.,   {de Blok} W.~J.~G.,  2008, \mn@doi
  [\apj] {10.1086/527543}, \href
  {https://ui.adsabs.harvard.edu/abs/2008ApJ...676..920K} {676, 920}

\bibitem[\protect\citeauthoryear{{Lancaster}, {Giovanetti}, {Mocz}, {Kahn},
  {Lisanti}  \& {Spergel}}{{Lancaster} et~al.}{2020}]{Lancaster2020}
{Lancaster} L.,  {Giovanetti} C.,  {Mocz} P.,  {Kahn} Y.,  {Lisanti} M.,
  {Spergel} D.~N.,  2020, \mn@doi [\jcap] {10.1088/1475-7516/2020/01/001},
  \href {https://ui.adsabs.harvard.edu/abs/2020JCAP...01..001L} {2020, 001}

\bibitem[\protect\citeauthoryear{{Lelli}, {McGaugh}  \& {Schombert}}{{Lelli}
  et~al.}{2016}]{Lelli2016}
{Lelli} F.,  {McGaugh} S.~S.,   {Schombert} J.~M.,  2016, \mn@doi [\aj]
  {10.3847/0004-6256/152/6/157}, \href
  {https://ui.adsabs.harvard.edu/abs/2016AJ....152..157L} {152, 157}

\bibitem[\protect\citeauthoryear{{Li}, {Vogelsberger}, {Marinacci}, {Sales}  \&
  {Torrey}}{{Li} et~al.}{2020}]{Li2020}
{Li} H.,  {Vogelsberger} M.,  {Marinacci} F.,  {Sales} L.~V.,   {Torrey} P.,
  2020, \mn@doi [\mnras] {10.1093/mnras/staa3122}, \href
  {https://ui.adsabs.harvard.edu/abs/2020MNRAS.499.5862L} {499, 5862}

\bibitem[\protect\citeauthoryear{{Ludlow}, {Schaye}, {Schaller}  \&
  {Richings}}{{Ludlow} et~al.}{2019a}]{Ludlow2019a}
{Ludlow} A.~D.,  {Schaye} J.,  {Schaller} M.,   {Richings} J.,  2019a, \mn@doi
  [\mnras] {10.1093/mnrasl/slz110}, \href
  {https://ui.adsabs.harvard.edu/abs/2019MNRAS.488L.123L} {488, L123}

\bibitem[\protect\citeauthoryear{{Ludlow}, {Schaye}  \& {Bower}}{{Ludlow}
  et~al.}{2019b}]{Ludlow2019b}
{Ludlow} A.~D.,  {Schaye} J.,   {Bower} R.,  2019b, \mn@doi [\mnras]
  {10.1093/mnras/stz1821}, \href
  {https://ui.adsabs.harvard.edu/abs/2019MNRAS.488.3663L} {488, 3663}

\bibitem[\protect\citeauthoryear{{Ludlow}, {Schaye}, {Schaller}  \&
  {Bower}}{{Ludlow} et~al.}{2020}]{Ludlow2020}
{Ludlow} A.~D.,  {Schaye} J.,  {Schaller} M.,   {Bower} R.,  2020, \mn@doi
  [\mnras] {10.1093/mnras/staa316}, \href
  {https://ui.adsabs.harvard.edu/abs/2020MNRAS.493.2926L} {493, 2926}

\bibitem[\protect\citeauthoryear{{Macci{\`o}}, {Stinson}, {Brook}, {Wadsley},
  {Couchman}, {Shen}, {Gibson}  \& {Quinn}}{{Macci{\`o}}
  et~al.}{2012}]{Maccio2012}
{Macci{\`o}} A.~V.,  {Stinson} G.,  {Brook} C.~B.,  {Wadsley} J.,  {Couchman}
  H.~M.~P.,  {Shen} S.,  {Gibson} B.~K.,   {Quinn} T.,  2012, \mn@doi [\apjl]
  {10.1088/2041-8205/744/1/L9}, \href
  {https://ui.adsabs.harvard.edu/abs/2012ApJ...744L...9M} {744, L9}

\bibitem[\protect\citeauthoryear{{Madau}, {Shen}  \& {Governato}}{{Madau}
  et~al.}{2014}]{madau2014}
{Madau} P.,  {Shen} S.,   {Governato} F.,  2014, \mn@doi [\apjl]
  {10.1088/2041-8205/789/1/L17}, \href
  {https://ui.adsabs.harvard.edu/abs/2014ApJ...789L..17M} {789, L17}

\bibitem[\protect\citeauthoryear{{Marasco}, {Oman}, {Navarro}, {Frenk}  \&
  {Oosterloo}}{{Marasco} et~al.}{2018}]{Marasco2018}
{Marasco} A.,  {Oman} K.~A.,  {Navarro} J.~F.,  {Frenk} C.~S.,   {Oosterloo}
  T.,  2018, \mn@doi [\mnras] {10.1093/mnras/sty354}, \href
  {https://ui.adsabs.harvard.edu/abs/2018MNRAS.476.2168M} {476, 2168}

\bibitem[\protect\citeauthoryear{{Marinacci}, {Pakmor}  \&
  {Springel}}{{Marinacci} et~al.}{2014}]{Marinacci2014}
{Marinacci} F.,  {Pakmor} R.,   {Springel} V.,  2014, \mn@doi [\mnras]
  {10.1093/mnras/stt2003}, \href
  {https://ui.adsabs.harvard.edu/abs/2014MNRAS.437.1750M} {437, 1750}

\bibitem[\protect\citeauthoryear{{Marinacci}, {Sales}, {Vogelsberger}, {Torrey}
   \& {Springel}}{{Marinacci} et~al.}{2019}]{Marinacci2019}
{Marinacci} F.,  {Sales} L.~V.,  {Vogelsberger} M.,  {Torrey} P.,   {Springel}
  V.,  2019, \mn@doi [\mnras] {10.1093/mnras/stz2391}, \href
  {https://ui.adsabs.harvard.edu/abs/2019MNRAS.489.4233M} {489, 4233}

\bibitem[\protect\citeauthoryear{{Mocz}, {Vogelsberger}, {Robles}, {Zavala},
  {Boylan-Kolchin}, {Fialkov}  \& {Hernquist}}{{Mocz} et~al.}{2017}]{Mocz2017}
{Mocz} P.,  {Vogelsberger} M.,  {Robles} V.~H.,  {Zavala} J.,  {Boylan-Kolchin}
  M.,  {Fialkov} A.,   {Hernquist} L.,  2017, \mn@doi [\mnras]
  {10.1093/mnras/stx1887}, \href
  {https://ui.adsabs.harvard.edu/abs/2017MNRAS.471.4559M} {471, 4559}

\bibitem[\protect\citeauthoryear{{Moore}}{{Moore}}{1994}]{Moore1994}
{Moore} B.,  1994, \mn@doi [\nat] {10.1038/370629a0}, \href
  {https://ui.adsabs.harvard.edu/abs/1994Natur.370..629M} {370, 629}

\bibitem[\protect\citeauthoryear{{Murray}, {Quataert}  \& {Thompson}}{{Murray}
  et~al.}{2005}]{Norman2005}
{Murray} N.,  {Quataert} E.,   {Thompson} T.~A.,  2005, \mn@doi [\apj]
  {10.1086/426067}, \href
  {https://ui.adsabs.harvard.edu/abs/2005ApJ...618..569M} {618, 569}

\bibitem[\protect\citeauthoryear{{Navarro}, {Eke}  \& {Frenk}}{{Navarro}
  et~al.}{1996a}]{navarro1996}
{Navarro} J.~F.,  {Eke} V.~R.,   {Frenk} C.~S.,  1996a, \mn@doi [\mnras]
  {10.1093/mnras/283.3.L72}, \href
  {https://ui.adsabs.harvard.edu/abs/1996MNRAS.283L..72N} {283, L72}

\bibitem[\protect\citeauthoryear{{Navarro}, {Frenk}  \& {White}}{{Navarro}
  et~al.}{1996b}]{navarro1996CDMhalos}
{Navarro} J.~F.,  {Frenk} C.~S.,   {White} S. D.~M.,  1996b, \mn@doi [\apj]
  {10.1086/177173}, \href
  {https://ui.adsabs.harvard.edu/abs/1996ApJ...462..563N} {462, 563}

\bibitem[\protect\citeauthoryear{{Oh}, {de Blok}, {Brinks}, {Walter}  \&
  {Kennicutt}}{{Oh} et~al.}{2011}]{Oh2011}
{Oh} S.-H.,  {de Blok} W.~J.~G.,  {Brinks} E.,  {Walter} F.,   {Kennicutt}
  Robert~C. J.,  2011, \mn@doi [\aj] {10.1088/0004-6256/141/6/193}, \href
  {https://ui.adsabs.harvard.edu/abs/2011AJ....141..193O} {141, 193}

\bibitem[\protect\citeauthoryear{{Oh} et~al.,}{{Oh} et~al.}{2015}]{Oh2015}
{Oh} S.-H.,  et~al., 2015, \mn@doi [\aj] {10.1088/0004-6256/149/6/180}, \href
  {https://ui.adsabs.harvard.edu/abs/2015AJ....149..180O} {149, 180}

\bibitem[\protect\citeauthoryear{{Oman} et~al.,}{{Oman}
  et~al.}{2015}]{Oman2015}
{Oman} K.~A.,  et~al., 2015, \mn@doi [\mnras] {10.1093/mnras/stv1504}, \href
  {https://ui.adsabs.harvard.edu/abs/2015MNRAS.452.3650O} {452, 3650}

\bibitem[\protect\citeauthoryear{{Oman}, {Marasco}, {Navarro}, {Frenk},
  {Schaye}  \& {Ben{\'\i}tez-Llambay}}{{Oman} et~al.}{2019}]{Oman2019}
{Oman} K.~A.,  {Marasco} A.,  {Navarro} J.~F.,  {Frenk} C.~S.,  {Schaye} J.,
  {Ben{\'\i}tez-Llambay} A.,  2019, \mn@doi [\mnras] {10.1093/mnras/sty2687},
  \href {https://ui.adsabs.harvard.edu/abs/2019MNRAS.482..821O} {482, 821}

\bibitem[\protect\citeauthoryear{{Padoan}, {Haugb{\o}lle}  \&
  {Nordlund}}{{Padoan} et~al.}{2012}]{Padoan2012}
{Padoan} P.,  {Haugb{\o}lle} T.,   {Nordlund} {\r{A}}.,  2012, \mn@doi [\apjl]
  {10.1088/2041-8205/759/2/L27}, \href
  {https://ui.adsabs.harvard.edu/abs/2012ApJ...759L..27P} {759, L27}

\bibitem[\protect\citeauthoryear{{Pillepich} et~al.,}{{Pillepich}
  et~al.}{2018}]{Pillepich2018}
{Pillepich} A.,  et~al., 2018, \mn@doi [\mnras] {10.1093/mnras/stx2656}, \href
  {https://ui.adsabs.harvard.edu/abs/2018MNRAS.473.4077P} {473, 4077}

\bibitem[\protect\citeauthoryear{{Pontzen} \& {Governato}}{{Pontzen} \&
  {Governato}}{2012}]{pontzengovernato2012}
{Pontzen} A.,  {Governato} F.,  2012, \mn@doi [\mnras]
  {10.1111/j.1365-2966.2012.20571.x}, \href
  {https://ui.adsabs.harvard.edu/abs/2012MNRAS.421.3464P} {421, 3464}

\bibitem[\protect\citeauthoryear{{Rahmati}, {Pawlik}, {Rai{\v{c}}evi{\'c}}  \&
  {Schaye}}{{Rahmati} et~al.}{2013}]{rahmati2013}
{Rahmati} A.,  {Pawlik} A.~H.,  {Rai{\v{c}}evi{\'c}} M.,   {Schaye} J.,  2013,
  \mn@doi [\mnras] {10.1093/mnras/stt066}, \href
  {https://ui.adsabs.harvard.edu/abs/2013MNRAS.430.2427R} {430, 2427}

\bibitem[\protect\citeauthoryear{{Read} \& {Gilmore}}{{Read} \&
  {Gilmore}}{2005}]{ReadGilmore2005}
{Read} J.~I.,  {Gilmore} G.,  2005, \mn@doi [\mnras]
  {10.1111/j.1365-2966.2004.08424.x}, \href
  {https://ui.adsabs.harvard.edu/abs/2005MNRAS.356..107R} {356, 107}

\bibitem[\protect\citeauthoryear{{Read}, {Iorio}, {Agertz}  \&
  {Fraternali}}{{Read} et~al.}{2016}]{read2016rc}
{Read} J.~I.,  {Iorio} G.,  {Agertz} O.,   {Fraternali} F.,  2016, \mn@doi
  [\mnras] {10.1093/mnras/stw1876}, \href
  {https://ui.adsabs.harvard.edu/abs/2016MNRAS.462.3628R} {462, 3628}

\bibitem[\protect\citeauthoryear{{Ren}, {Kwa}, {Kaplinghat}  \& {Yu}}{{Ren}
  et~al.}{2019}]{Ren2019}
{Ren} T.,  {Kwa} A.,  {Kaplinghat} M.,   {Yu} H.-B.,  2019, \mn@doi [Physical
  Review X] {10.1103/PhysRevX.9.031020}, \href
  {https://ui.adsabs.harvard.edu/abs/2019PhRvX...9c1020R} {9, 031020}

\bibitem[\protect\citeauthoryear{{Robles} et~al.,}{{Robles}
  et~al.}{2017}]{Robles2017}
{Robles} V.~H.,  et~al., 2017, \mn@doi [\mnras] {10.1093/mnras/stx2253}, \href
  {https://ui.adsabs.harvard.edu/abs/2017MNRAS.472.2945R} {472, 2945}

\bibitem[\protect\citeauthoryear{{Rocha}, {Peter}, {Bullock}, {Kaplinghat},
  {Garrison-Kimmel}, {O{\~n}orbe}  \& {Moustakas}}{{Rocha}
  et~al.}{2013}]{Rocha2013}
{Rocha} M.,  {Peter} A. H.~G.,  {Bullock} J.~S.,  {Kaplinghat} M.,
  {Garrison-Kimmel} S.,  {O{\~n}orbe} J.,   {Moustakas} L.~A.,  2013, \mn@doi
  [\mnras] {10.1093/mnras/sts514}, \href
  {https://ui.adsabs.harvard.edu/abs/2013MNRAS.430...81R} {430, 81}

\bibitem[\protect\citeauthoryear{{Rogstad}, {Lockhart}  \& {Wright}}{{Rogstad}
  et~al.}{1974}]{Rogstad1974}
{Rogstad} D.~H.,  {Lockhart} I.~A.,   {Wright} M.~C.~H.,  1974, \mn@doi [\apj]
  {10.1086/153164}, \href
  {https://ui.adsabs.harvard.edu/abs/1974ApJ...193..309R} {193, 309}

\bibitem[\protect\citeauthoryear{{Sales}, {Marinacci}, {Springel}  \&
  {Petkova}}{{Sales} et~al.}{2014}]{Sales2014}
{Sales} L.~V.,  {Marinacci} F.,  {Springel} V.,   {Petkova} M.,  2014, \mn@doi
  [\mnras] {10.1093/mnras/stu155}, \href
  {https://ui.adsabs.harvard.edu/abs/2014MNRAS.439.2990S} {439, 2990}

\bibitem[\protect\citeauthoryear{{Santos-Santos}, {Di Cintio}, {Brook},
  {Macci{\`o}}, {Dutton}  \& {Dom{\'\i}nguez-Tenreiro}}{{Santos-Santos}
  et~al.}{2018}]{Santos-Santos2018}
{Santos-Santos} I.~M.,  {Di Cintio} A.,  {Brook} C.~B.,  {Macci{\`o}} A.,
  {Dutton} A.,   {Dom{\'\i}nguez-Tenreiro} R.,  2018, \mn@doi [\mnras]
  {10.1093/mnras/stx2660}, \href
  {https://ui.adsabs.harvard.edu/abs/2018MNRAS.473.4392S} {473, 4392}

\bibitem[\protect\citeauthoryear{{Santos-Santos} et~al.,}{{Santos-Santos}
  et~al.}{2020}]{Santos-Santos2020}
{Santos-Santos} I. M.~E.,  et~al., 2020, \mn@doi [\mnras]
  {10.1093/mnras/staa1072}, \href
  {https://ui.adsabs.harvard.edu/abs/2020MNRAS.495...58S} {495, 58}

\bibitem[\protect\citeauthoryear{{Sawala} et~al.,}{{Sawala}
  et~al.}{2016}]{sawala2016}
{Sawala} T.,  et~al., 2016, \mn@doi [\mnras] {10.1093/mnras/stw145}, \href
  {https://ui.adsabs.harvard.edu/abs/2016MNRAS.457.1931S} {457, 1931}

\bibitem[\protect\citeauthoryear{{Schaller} et~al.,}{{Schaller}
  et~al.}{2015}]{Schaller2015}
{Schaller} M.,  et~al., 2015, \mn@doi [\mnras] {10.1093/mnras/stv1067}, \href
  {https://ui.adsabs.harvard.edu/abs/2015MNRAS.451.1247S} {451, 1247}

\bibitem[\protect\citeauthoryear{{Schaye} et~al.,}{{Schaye}
  et~al.}{2015}]{Schaye2015}
{Schaye} J.,  et~al., 2015, \mn@doi [\mnras] {10.1093/mnras/stu2058}, \href
  {https://ui.adsabs.harvard.edu/abs/2015MNRAS.446..521S} {446, 521}

\bibitem[\protect\citeauthoryear{{Semenov}, {Kravtsov}  \& {Gnedin}}{{Semenov}
  et~al.}{2016}]{Semenov2016}
{Semenov} V.~A.,  {Kravtsov} A.~V.,   {Gnedin} N.~Y.,  2016, \mn@doi [\apj]
  {10.3847/0004-637X/826/2/200}, \href
  {https://ui.adsabs.harvard.edu/abs/2016ApJ...826..200S} {826, 200}

\bibitem[\protect\citeauthoryear{{Smith}, {Safranek-Shrader}, {Bromm}  \&
  {Milosavljevi{\'c}}}{{Smith} et~al.}{2015}]{Smith2015}
{Smith} A.,  {Safranek-Shrader} C.,  {Bromm} V.,   {Milosavljevi{\'c}} M.,
  2015, \mn@doi [\mnras] {10.1093/mnras/stv565}, \href
  {https://ui.adsabs.harvard.edu/abs/2015MNRAS.449.4336S} {449, 4336}

\bibitem[\protect\citeauthoryear{{Smith}, {Bromm}  \& {Loeb}}{{Smith}
  et~al.}{2017}]{Smith2017}
{Smith} A.,  {Bromm} V.,   {Loeb} A.,  2017, \mn@doi [\mnras]
  {10.1093/mnras/stw2591}, \href
  {https://ui.adsabs.harvard.edu/abs/2017MNRAS.464.2963S} {464, 2963}

\bibitem[\protect\citeauthoryear{{Smith}, {Sijacki}  \& {Shen}}{{Smith}
  et~al.}{2018}]{smith2018}
{Smith} M.~C.,  {Sijacki} D.,   {Shen} S.,  2018, \mn@doi [\mnras]
  {10.1093/mnras/sty994}, \href
  {https://ui.adsabs.harvard.edu/abs/2018MNRAS.478..302S} {478, 302}

\bibitem[\protect\citeauthoryear{{Spergel} \& {Steinhardt}}{{Spergel} \&
  {Steinhardt}}{2000}]{Spergel2000}
{Spergel} D.~N.,  {Steinhardt} P.~J.,  2000, \mn@doi [\prl]
  {10.1103/PhysRevLett.84.3760}, \href
  {https://ui.adsabs.harvard.edu/abs/2000PhRvL..84.3760S} {84, 3760}

\bibitem[\protect\citeauthoryear{{Springel}}{{Springel}}{2010}]{arepo}
{Springel} V.,  2010, \mn@doi [\mnras] {10.1111/j.1365-2966.2009.15715.x},
  \href {https://ui.adsabs.harvard.edu/abs/2010MNRAS.401..791S} {401, 791}

\bibitem[\protect\citeauthoryear{{Springel} \& {Hernquist}}{{Springel} \&
  {Hernquist}}{2003}]{SpringelHernquist2003}
{Springel} V.,  {Hernquist} L.,  2003, \mn@doi [\mnras]
  {10.1046/j.1365-8711.2003.06206.x}, \href
  {https://ui.adsabs.harvard.edu/abs/2003MNRAS.339..289S} {339, 289}

\bibitem[\protect\citeauthoryear{{Springel}, {Di Matteo}  \&
  {Hernquist}}{{Springel} et~al.}{2005}]{Springel2005}
{Springel} V.,  {Di Matteo} T.,   {Hernquist} L.,  2005, \mn@doi [\mnras]
  {10.1111/j.1365-2966.2005.09238.x}, \href
  {https://ui.adsabs.harvard.edu/abs/2005MNRAS.361..776S} {361, 776}

\bibitem[\protect\citeauthoryear{{Springel} et~al.,}{{Springel}
  et~al.}{2008}]{Springel2008}
{Springel} V.,  et~al., 2008, \mn@doi [\mnras]
  {10.1111/j.1365-2966.2008.14066.x}, \href
  {https://ui.adsabs.harvard.edu/abs/2008MNRAS.391.1685S} {391, 1685}

\bibitem[\protect\citeauthoryear{{Stinson}, {Seth}, {Katz}, {Wadsley},
  {Governato}  \& {Quinn}}{{Stinson} et~al.}{2006}]{Stinson2006}
{Stinson} G.,  {Seth} A.,  {Katz} N.,  {Wadsley} J.,  {Governato} F.,   {Quinn}
  T.,  2006, \mn@doi [\mnras] {10.1111/j.1365-2966.2006.11097.x}, \href
  {https://ui.adsabs.harvard.edu/abs/2006MNRAS.373.1074S} {373, 1074}

\bibitem[\protect\citeauthoryear{{Teyssier}, {Pontzen}, {Dubois}  \&
  {Read}}{{Teyssier} et~al.}{2013}]{Teyssier2013}
{Teyssier} R.,  {Pontzen} A.,  {Dubois} Y.,   {Read} J.~I.,  2013, \mn@doi
  [\mnras] {10.1093/mnras/sts563}, \href
  {https://ui.adsabs.harvard.edu/abs/2013MNRAS.429.3068T} {429, 3068}

\bibitem[\protect\citeauthoryear{{Thielemann} et~al.,}{{Thielemann}
  et~al.}{2003}]{thielemann2003}
{Thielemann} F.~K.,  et~al., 2003, in {Hillebrandt} W.,  {Leibundgut} B.,  eds,
  From Twilight to Highlight: The Physics of Supernovae. p.~331,
  \mn@doi{10.1007/10828549_46}

\bibitem[\protect\citeauthoryear{{Tollet} et~al.,}{{Tollet}
  et~al.}{2016}]{tollet2016}
{Tollet} E.,  et~al., 2016, \mn@doi [\mnras] {10.1093/mnras/stv2856}, \href
  {https://ui.adsabs.harvard.edu/abs/2016MNRAS.456.3542T} {456, 3542}

\bibitem[\protect\citeauthoryear{{Tulin} \& {Yu}}{{Tulin} \&
  {Yu}}{2018}]{TulinYu2018}
{Tulin} S.,  {Yu} H.-B.,  2018, \mn@doi [\physrep]
  {10.1016/j.physrep.2017.11.004}, \href
  {https://ui.adsabs.harvard.edu/abs/2018PhR...730....1T} {730, 1}

\bibitem[\protect\citeauthoryear{{Vogelsberger}, {Zavala}  \&
  {Loeb}}{{Vogelsberger} et~al.}{2012}]{Vogelsberger2012}
{Vogelsberger} M.,  {Zavala} J.,   {Loeb} A.,  2012, \mn@doi [\mnras]
  {10.1111/j.1365-2966.2012.21182.x}, \href
  {https://ui.adsabs.harvard.edu/abs/2012MNRAS.423.3740V} {423, 3740}

\bibitem[\protect\citeauthoryear{{Vogelsberger}, {Genel}, {Sijacki}, {Torrey},
  {Springel}  \& {Hernquist}}{{Vogelsberger} et~al.}{2013}]{v13feedback}
{Vogelsberger} M.,  {Genel} S.,  {Sijacki} D.,  {Torrey} P.,  {Springel} V.,
  {Hernquist} L.,  2013, \mn@doi [\mnras] {10.1093/mnras/stt1789}, \href
  {https://ui.adsabs.harvard.edu/abs/2013MNRAS.436.3031V} {436, 3031}

\bibitem[\protect\citeauthoryear{{Vogelsberger} et~al.,}{{Vogelsberger}
  et~al.}{2014a}]{v14illustris}
{Vogelsberger} M.,  et~al., 2014a, \mn@doi [\mnras] {10.1093/mnras/stu1536},
  \href {https://ui.adsabs.harvard.edu/abs/2014MNRAS.444.1518V} {444, 1518}

\bibitem[\protect\citeauthoryear{{Vogelsberger}, {Zavala}, {Simpson}  \&
  {Jenkins}}{{Vogelsberger} et~al.}{2014b}]{v14dwarfs}
{Vogelsberger} M.,  {Zavala} J.,  {Simpson} C.,   {Jenkins} A.,  2014b, \mn@doi
  [\mnras] {10.1093/mnras/stu1713}, \href
  {https://ui.adsabs.harvard.edu/abs/2014MNRAS.444.3684V} {444, 3684}

\bibitem[\protect\citeauthoryear{{Vogelsberger} et~al.,}{{Vogelsberger}
  et~al.}{2014c}]{v14nature}
{Vogelsberger} M.,  et~al., 2014c, \mn@doi [\nat] {10.1038/nature13316}, \href
  {https://ui.adsabs.harvard.edu/abs/2014Natur.509..177V} {509, 177}

\bibitem[\protect\citeauthoryear{{Vogelsberger}, {Zavala}, {Schutz}  \&
  {Slatyer}}{{Vogelsberger} et~al.}{2019}]{Vogelsberger2019}
{Vogelsberger} M.,  {Zavala} J.,  {Schutz} K.,   {Slatyer} T.~R.,  2019,
  \mn@doi [\mnras] {10.1093/mnras/stz340}, \href
  {https://ui.adsabs.harvard.edu/abs/2019MNRAS.484.5437V} {484, 5437}

\bibitem[\protect\citeauthoryear{{Vogelsberger}, {Marinacci}, {Torrey}  \&
  {Puchwein}}{{Vogelsberger} et~al.}{2020}]{Vogelsberger2020}
{Vogelsberger} M.,  {Marinacci} F.,  {Torrey} P.,   {Puchwein} E.,  2020,
  \mn@doi [Nature Reviews Physics] {10.1038/s42254-019-0127-2}, \href
  {https://ui.adsabs.harvard.edu/abs/2020NatRP...2...42V} {2, 42}

\bibitem[\protect\citeauthoryear{{Wadsley}, {Stadel}  \& {Quinn}}{{Wadsley}
  et~al.}{2004}]{Wadsley2004}
{Wadsley} J.~W.,  {Stadel} J.,   {Quinn} T.,  2004, \mn@doi [\na]
  {10.1016/j.newast.2003.08.004}, \href
  {https://ui.adsabs.harvard.edu/abs/2004NewA....9..137W} {9, 137}

\bibitem[\protect\citeauthoryear{{Walker} \& {Pe{\~n}arrubia}}{{Walker} \&
  {Pe{\~n}arrubia}}{2011}]{Walker2011}
{Walker} M.~G.,  {Pe{\~n}arrubia} J.,  2011, \mn@doi [\apj]
  {10.1088/0004-637X/742/1/20}, \href
  {https://ui.adsabs.harvard.edu/abs/2011ApJ...742...20W} {742, 20}

\bibitem[\protect\citeauthoryear{{Wang}, {Dutton}, {Stinson}, {Macci{\`o}},
  {Penzo}, {Kang}, {Keller}  \& {Wadsley}}{{Wang} et~al.}{2015}]{wang2015nihao}
{Wang} L.,  {Dutton} A.~A.,  {Stinson} G.~S.,  {Macci{\`o}} A.~V.,  {Penzo} C.,
   {Kang} X.,  {Keller} B.~W.,   {Wadsley} J.,  2015, \mn@doi [\mnras]
  {10.1093/mnras/stv1937}, \href
  {https://ui.adsabs.harvard.edu/abs/2015MNRAS.454...83W} {454, 83}

\bibitem[\protect\citeauthoryear{{Weinberger}, {Springel}  \&
  {Pakmor}}{{Weinberger} et~al.}{2020}]{Weinberger2020}
{Weinberger} R.,  {Springel} V.,   {Pakmor} R.,  2020, \mn@doi [\apjs]
  {10.3847/1538-4365/ab908c}, \href
  {https://ui.adsabs.harvard.edu/abs/2020ApJS..248...32W} {248, 32}

\bibitem[\protect\citeauthoryear{{Wetzel}, {Hopkins}, {Kim},
  {Faucher-Gigu{\`e}re}, {Kere{\v{s}}}  \& {Quataert}}{{Wetzel}
  et~al.}{2016}]{wetzel2016LATTE}
{Wetzel} A.~R.,  {Hopkins} P.~F.,  {Kim} J.-h.,  {Faucher-Gigu{\`e}re} C.-A.,
  {Kere{\v{s}}} D.,   {Quataert} E.,  2016, \mn@doi [\apj]
  {10.3847/2041-8205/827/2/L23}, \href
  {https://ui.adsabs.harvard.edu/abs/2016ApJ...827L..23W} {827, L23}

\bibitem[\protect\citeauthoryear{{White} \& {Rees}}{{White} \&
  {Rees}}{1978}]{white1978}
{White} S.~D.~M.,  {Rees} M.~J.,  1978, \mn@doi [\mnras]
  {10.1093/mnras/183.3.341}, \href
  {https://ui.adsabs.harvard.edu/abs/1978MNRAS.183..341W} {183, 341}

\bibitem[\protect\citeauthoryear{{Yoshida}, {Springel}, {White}  \&
  {Tormen}}{{Yoshida} et~al.}{2000}]{Yoshida2000}
{Yoshida} N.,  {Springel} V.,  {White} S. D.~M.,   {Tormen} G.,  2000, \mn@doi
  [\apjl] {10.1086/317306}, \href
  {https://ui.adsabs.harvard.edu/abs/2000ApJ...544L..87Y} {544, L87}

\bibitem[\protect\citeauthoryear{{Zavala}, {Lovell}, {Vogelsberger}  \&
  {Burger}}{{Zavala} et~al.}{2019}]{Zavala2019}
{Zavala} J.,  {Lovell} M.~R.,  {Vogelsberger} M.,   {Burger} J.~D.,  2019,
  \mn@doi [\prd] {10.1103/PhysRevD.100.063007}, \href
  {https://ui.adsabs.harvard.edu/abs/2019PhRvD.100f3007Z} {100, 063007}

\bibitem[\protect\citeauthoryear{{Zolotov} et~al.,}{{Zolotov}
  et~al.}{2012}]{Zolotov2012}
{Zolotov} A.,  et~al., 2012, \mn@doi [\apj] {10.1088/0004-637X/761/1/71}, \href
  {https://ui.adsabs.harvard.edu/abs/2012ApJ...761...71Z} {761, 71}

\bibitem[\protect\citeauthoryear{{de Blok}, {McGaugh}, {Bosma}  \& {Rubin}}{{de
  Blok} et~al.}{2001}]{deblok2001}
{de Blok} W.~J.~G.,  {McGaugh} S.~S.,  {Bosma} A.,   {Rubin} V.~C.,  2001,
  \mn@doi [\apjl] {10.1086/320262}, \href
  {https://ui.adsabs.harvard.edu/abs/2001ApJ...552L..23D} {552, L23}

\bibitem[\protect\citeauthoryear{{de Blok}, {Walter}, {Brinks}, {Trachternach},
  {Oh}  \& {Kennicutt}}{{de Blok} et~al.}{2008}]{deBlok2008}
{de Blok} W.~J.~G.,  {Walter} F.,  {Brinks} E.,  {Trachternach} C.,  {Oh}
  S.~H.,   {Kennicutt} R.~C. J.,  2008, \mn@doi [\aj]
  {10.1088/0004-6256/136/6/2648}, \href
  {https://ui.adsabs.harvard.edu/abs/2008AJ....136.2648D} {136, 2648}

\bibitem[\protect\citeauthoryear{{van den Bosch} \& {Ogiya}}{{van den Bosch} \&
  {Ogiya}}{2018}]{vandenBosch2018}
{van den Bosch} F.~C.,  {Ogiya} G.,  2018, \mn@doi [\mnras]
  {10.1093/mnras/sty084}, \href
  {https://ui.adsabs.harvard.edu/abs/2018MNRAS.475.4066V} {475, 4066}

\makeatother
\end{thebibliography}




\appendix

\section{Additional Material}
\label{sec:appendix}

\begin{figure}
    \centering
    \includegraphics[width=0.45\textwidth]{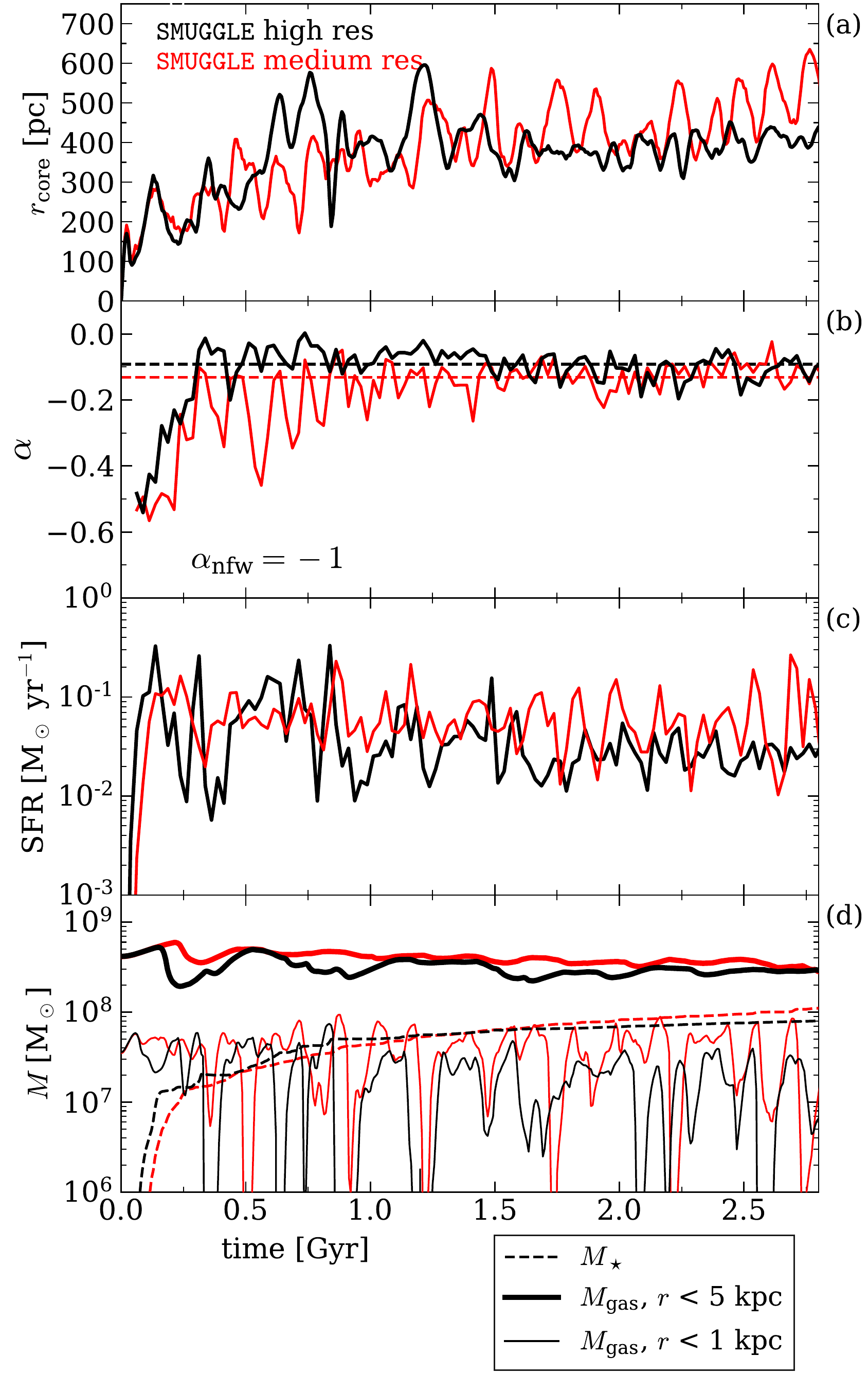}
    \caption{Comparison between the fiducial \texttt{SMUGGLE} model ran at our presented resolution of $m_\text{bary} = 850$ \msun~(high res), and at a $10\times$ lower resolution of $m_\text{bary} = 8500$ \msun~(medium res). }
    \label{fig:res_converge}
\end{figure}

To demonstrate the resolution convergence of the \texttt{SMUGGLE} model, Figure \ref{fig:res_converge} shows our standard high resolution run with $m_\text{bary} = 850$ \msun~compared to a run at $10$ times lower resolution ($m_\text{bary} = 8500$ \msun). We find good agreement in all measured quantities. The lower resolution run demonstrates marginally less bursty star formation rates. We also find stronger time dependence in the measured core radius and power law slope in the lower resolution run, but good agreement in values for these quantities across the run time. A notable similarity between the runs is their rapid fluctuations in gas mass within $1$ kpc. 

\begin{figure}
    \centering
    \includegraphics[width=0.45\textwidth]{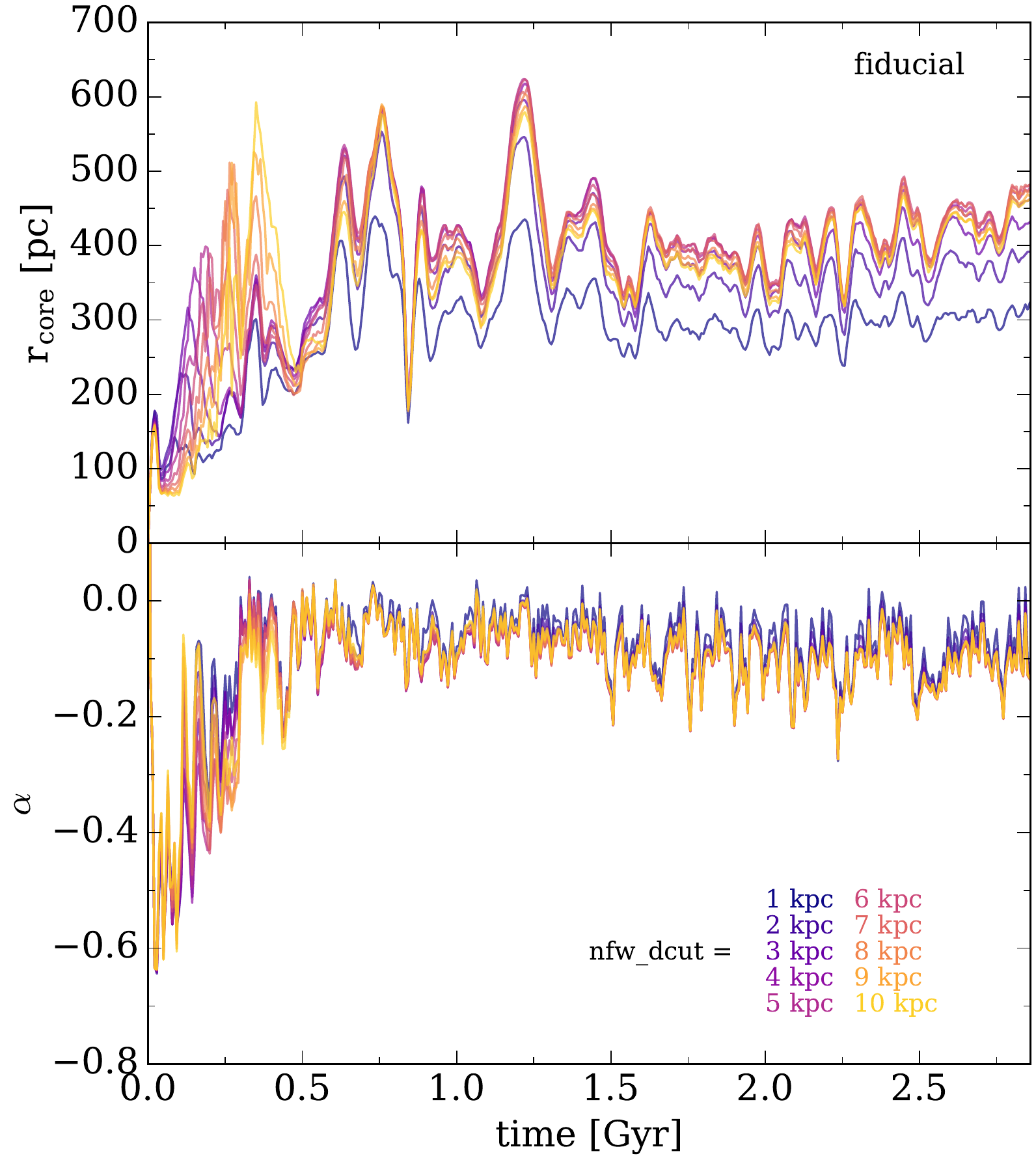}
    \caption{Comparison of the core radius and slope measured using NFW reference profiles fit to $r >$ nfw\_dcut, as listed on the figure. We find that our measurements are robust to choice of this parameter. }
    \label{fig:rcore_slope_compare_nfwdcut}
\end{figure}

Figure \ref{fig:rcore_slope_compare_nfwdcut} shows the measured values of core radius and inner power law slope for the fiducial \texttt{SMUGGLE} model with different values of radial cutoff for the fitted NFW profile (see Section \ref{sec:coresize}). We find that the measured values of \rcore~and inner slope $\alpha$ are robust to choice of NFW radial fitting cutoff in the range $1 - 10$ kpc. 

\begin{figure*}
    \centering
    \includegraphics[width=\textwidth]{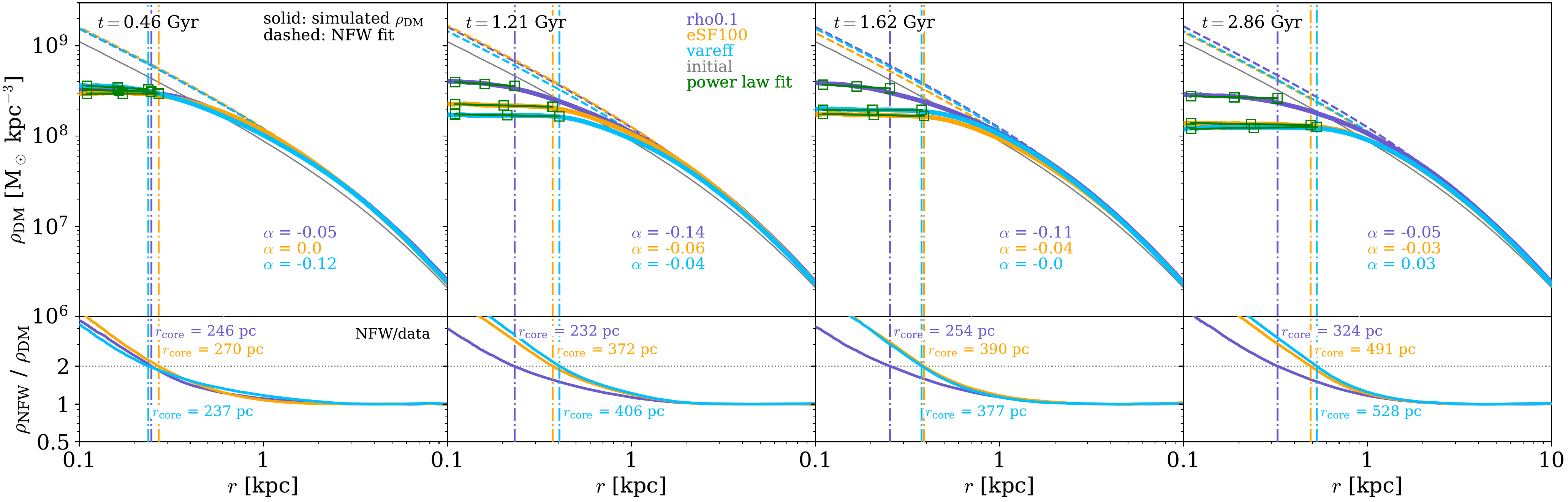}
    \caption{Dark matter density profiles at select times for the runs with variations in the ISM model of \texttt{SMUGGLE}. Dashed lines indicate NFW profiles fit to the run of the corresponding color. Power law fits are shown as dark green square, and the core radius is indicated with a vertical line. }
    \label{fig:fourrho_varISM}
\end{figure*}

Figure \ref{fig:fourrho_varISM} depicts density profiles for the three \texttt{SMUGGLE} variations we explore in Section \ref{sec:smuggle_vars}, similar to Figure \ref{fig:rho_panel}. Each panel shows a different timestamp of the simulation, including the final snapshot in the rightmost panel. The bottom panels, as in Figure \ref{fig:rho_panel}, show the ratio of densities of the fitted NFW profile to the physical DM density achieved in the simulation. The NFW profile is fitted to $r > 3$ kpc, and the core radius is chosen as the radial distance where $\rho_\text{NFW} = 2 \rho_\text{dm}$. The core radius is depicted as a vertical dash-dot line, and its value at each timestamp is listed on the bottom panels. We find that \texttt{eSF100} and \texttt{vareff}, models with increased local star formation efficiency, demonstrate large, flattened core. Our run with density threshold of $\rho = 0.1$ cm$^{-3}$ (\texttt{rho0.1}), forms a core that is somewhat smaller than those of the high efficiency runs, though it is still consistent with the fiducial \texttt{SMUGGLE} model, and clearly distinct from the lack of core found in \texttt{SH03}, as discussed in Section \ref{sec:smuggle_vars}.

\begin{figure}
    \centering
    \includegraphics[width=0.45\textwidth]{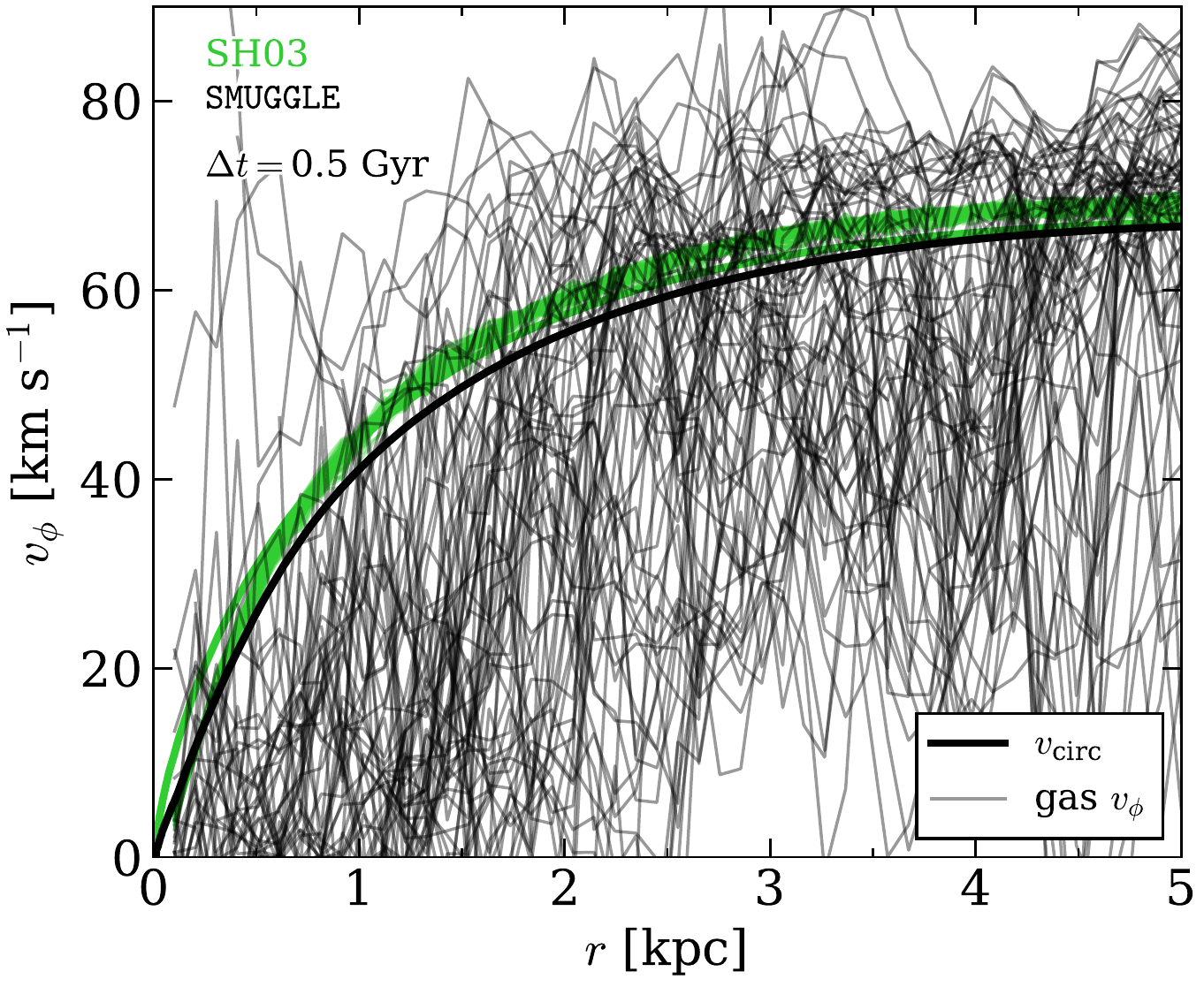}
    \caption{Gas rotational velocity profiles of the fiducial \texttt{SMUGGLE} model for each snapshot in the last 0.5 Gyr of its run time. Realistic modeling of ISM physics produces large variation in the rotational component of gas within the galaxy, leading to a large diversity of rotation curves.}
    \label{fig:vphi_all}
\end{figure}

We plot the rotational velocity profiles of gas for both the fiducial \texttt{SMUGGLE} model and SH03 \citep[][]{SpringelHernquist2003} for each snapshot of the final $0.5$ Gyr of run time in Figure \ref{fig:vphi_all}. The median gas $v_\phi$ profile for all models is shown in Figure \ref{fig:vphi}. We plot these individual profiles to explicitly show the variety of shapes that can be produced through feedback effects on the gas in \texttt{SMUGGLE}, and the uniformity of profiles achieved in SH03. We find that a handful of profiles (no more than 10 per cent) demonstrate steeply rising inner velocity profiles, suggesting that baryonic feedback can account for some of the observed diversity of rotation curves. Overall, we find rotational velocity profiles that tend to underestimate the circular velocity (thicker lines). These rotational profiles can also demonstrate strong fluctuations with radius, as well as time, indicating a gaseous component that is in a constant state of flux.


\bsp	
\label{lastpage}
\end{document}